\documentclass{aa}
\usepackage{keyval}
\usepackage{graphicx}
\usepackage{txfonts}
\usepackage{hyperref}
\usepackage{academicons}

\usepackage{natbib,twoopt}
\usepackage[dvipsnames]{xcolor}

\newcommand{\Gaia}{{\it Gaia}}

\newcommand{\gspspec}{{\it GSP-Spec}}

\newcommand{\T}{$T_{\rm eff}$}
\newcommand{\g}{log($g$)}

\newcommand{\vunits}{km/s}
\newcommand{\meta}{[M/H]}
\newcommand{\alfa}{$\alpha$}
\newcommand{\AF}{[\alfa/Fe]}

\newcommand{\Vrad}{$V_{\rm Rad}$}
\newcommand{\Vmi}{$V_{\text{micro}}$}
\newcommand{\Vma}{$V_{\text{macro}}$}
\newcommand{\Vsini}{$V_{\text{sin $i$}}$}
\newcommand{\Vb}{$V_{\rm broad}$}

\newcommand{\SNR}{$S/N$}

\usepackage{listings}
\definecolor{dkgreen}{rgb}{0,0.6,0}
\definecolor{gray}{rgb}{0.5,0.5,0.5}
\definecolor{mauve}{rgb}{0.58,0,0.82}

\begin{document} 

   \title{The AMBRE Project: Line-broadening and stellar rotation of  ESO/FEROS archived spectra}

   \author{   
          F. Bado \inst{1, 2}\thanks{Corresponding author P. de Laverny}
          \and
          P. de Laverny \inst{1}
          \and
          Z. Kam\inst{2}  
          \and
          A. Recio-Blanco\inst{1}
          \and
          P.A. Palicio\inst{1}
          \and
          J. Koulidiati\inst{2}            
          } 

  \institute{Université Côte d'Azur, Observatoire de la Côte d'Azur, CNRS, Laboratoire Lagrange, Bd de l'Observatoire, CS 34229, 06304 Nice cedex 4, France.
    \email{fabricebado726@gmail.com}
    \and
      Université Jospeh KI-ZERBO, Département de Physique, Laboratoire de Physique et de Chimie de l'Environnement, 03 BP 7021, Ouagadougou, Burkina Faso.
   }

   \date{Received 2025, September ?? ; accepted }
   
   \abstract
   {Stellar rotation is a fundamental parameter in studies of the formation and evolution of stars. However,  large homogeneous catalogues of rotational velocities derived from high-resolution stellar spectra are still lacking.}
   {The main objective of this work is to determine the line-broadening parameter (\Vb), which is a proxy for the  stellar rotational velocity, in a large sample of FGKM stars based on their ESO/FEROS spectra. All these stars were previously parameterised by the AMBRE Project.}
   {The line-broadening parameter was estimated by cross-correlating the FEROS spectra with binary masks, specifically chosen on the basis of the AMBRE stellar parameters. This methodology also relies on a specific calibration of a coupling constant between the rotational velocity and the width of the cross-correlation function. This fundamental step was performed by adopting the AMBRE grid of synthetic spectra. The derived \Vb\ were then validated using data from the literature, ground-based spectroscopic surveys, and Gaia/RVS.}
   {After analysing more than 5,000 FEROS spectra (including repeated spectra for several stars), we obtained the line-broadening coefficients for 2,584 stars covering the FGKM spectral types, any stellar gravity, and metallicities between the metal-poor up to
   sub-solar regimes. The mean \Vb\ relative uncertainty of this sample was found to be smaller than 8\%. As expected, most stars were found to be slow rotators (below a few \vunits), in particular, cool dwarfs and giants. However, several hot dwarfs and high-luminosity stars with high-\Vb\ rates were identified, most of them not previously classified as fast rotators and/or affected by large macro-turbulent effects before the present work.}
   {The measured rotational
broadening values are of high-quality and verified on the basis of literature comparisons. We publicly provide this catalogue of line-broadening parameters, including stellar atmospheric and quality parameters, for the analysed AMBRE/FEROS sources.}
   
   \maketitle

\section{Introduction}
\label{Sec:Intro}
The accurate determination of fundamental stellar parameters, including effective temperature, surface gravity, chemical composition, luminosity, mass, radius, and rotational velocity for large catalogues remains a significant challenge in astrophysics. However, recent advancements have been made, driven by large-scale spectroscopic surveys from the
ground as the \Gaia\ ESO Survey \citep{Gilmore2012}, AMBRE \citep[][see below]{deLaverny2013}, APOGEE \citep{Majewski2017}, and GALAH~\citep{De_Silva2015}
or from space, such as the European Space Agency (ESA) \Gaia\ mission~\citep{Gaia_Collaboration2016} through its \gspspec\ catalogue \citep{Recio2023}.
Previously considered as a lower important parameter, stellar rotation is now recognised as playing a key role; for instance, it is crucial with respect to tracing a star’s structure and evolution~\citep{Maeder2000}. It can also influence internal mixing, elemental diffusion, stellar winds, and magnetic field generation. For instance, horizontal branch stars in globular clusters are known to rotate at different rate, depending on their location on this branch \citep{Recio2002}.
Stellar rotation can also be considered as a chronometer, aiding in age determination through gyrochronology~\citep{Barnes2007, Angus2015}.
Rotational velocity distribution of stellar populations can also provide valuable insights into star formation processes and evolutionary paths~\citep{Zorec2012}. Finally, understanding stellar rotation can help improve chemical enrichment models \citep{Chiappini2003, Prantzos18, Prantzos20}. 

From the observational point of view, several studies have produced catalogues of stellar rotational velocities. First, we recall that the rotational velocity (\Vsini) is the product of the equatorial rotational velocity of the star by the sine of the inclination angle, $i$, between the star’s rotation axis and the observer’s line of sight. 
Then, it is worth noting that several of the published catalogues do not
directly report \Vsini\,, but instead a line-broadening parameter (\Vb) that is a good estimate of stellar rotation.
The absorption lines present in any stellar spectra are indeed broadened by the combination of several physical mechanisms inherent to stellar atmospheres: the projected rotational velocity (\Vsini), the macro-turbulence velocity (\Vma), and the micro-turbulence velocity (\Vmi). In most stars, \Vb~is dominated by \Vsini~as soon as it is larger than a few \vunits, but making a direct measurement of \Vsini\ from stellar spectra is not straightforward, since the other broadening mechanism have to be independently measured. Among the stellar rotation catalogues published before the era of large spectroscopic surveys, we could cite, for instance, the important works of \citet[and previous studies on cool stars by the same group]
{JRM2006} and \cite{GG2005}. Regarding stellar rotation in hotter stars,
we could cite \citet[IACOB project, OB-type stars]{Holgado22, Burgos25}, \citet[B-type dwarfs]{Bragan12}, or \citet[A-type stars]{Zorec12}.
Then, large spectroscopic surveys as APOGEE, GALAH, and the ESA \Gaia\ mission \citep{Abdurrouf2022, Buder2024, Fremat2023} published their own significant catalogues of stellar rotational rates. Most of these catalogues are detailed in Sect.~\ref{Sec:Catalog} since they have been adopted to validate our measurements in this work.
To complement such catalogues thanks to high-resolution spectra, 
the present work is aimed at measuring the spectral line broadening (\Vb) of the FEROS spectra, previously parameterised by the AMBRE Project. Because of the large parameter space covered, this new catalogue delivers constraints on the rotational rate of different stellar types.

This article is structured as follows. The observational data considered in this work are described in Sect.~\ref{Sec:data}. In Sect.~\ref{Sec:Method}, we present the  methodology adopted to measure the line-broadening velocity in individual
FEROS spectra. The AMBRE/FEROS catalogue of stars with measured \Vb\ values is presented in Sect.~\ref{Sec:Catalog} and compared to results from the  literature in Sect.~\ref{Sec:Valid}. Some rotational properties of the studied sample are discussed in Sect.~\ref{Sec:Discussion}, while we summarise our study in Sect.~\ref{Sec:Conclu}. 

\section{The AMBRE:FEROS data}
\label{Sec:data}
 The AMBRE Project consists of a stellar parameterisation and a Galactic Archaeology initiative led by the Observatoire de la Côte d'Azur \citep{deLaverny2013}. AMBRE has automatically parametrised a few $10^5$ high-resolution stellar spectra collected by instruments of the European Southern Observatory (ESO). The extracted parameters include
the radial velocities (\Vrad) and the main atmospheric parameters: \T, \g, \meta, and enrichment in $\alpha$-elements versus iron abundances, \AF\ \citep[see][for the analysis of FEROS, HARPS and UVES spectra, respectively]{worley2012, DePascale2014, worley2016}. Moreover, the individual chemical abundances of various elements as, for instance, sulphur and lead, were also extracted \citep{Perdigon2021, Contursi2024}. Finally,  AMBRE  is also among the first efforts undertaken towards the automatic parameterisation of the million of spectra collected by the European Space Agency (ESA) \Gaia\ mission \citep{Recio2023}.

Regarding the spectra analysed in the present work, they  were all collected with FEROS (Fiber-fed Extended Range Optical Spectrograph), an instrument mount on the MPG/ESO 2.2~m telescope. FEROS
covers the $\sim$350~nm to $\sim$920~nm spectral range with a spectral resolution of $R$ = 48\,000 \citep{Kaufer1999} and two wavelength samplings of 0.003 nm and 0.006 nm. Within AMBRE, 
\cite{worley2012}  analysed 21\,551 FEROS archived spectra, which correspond to 6\,285 
different stars. Among them, 30.2\%  ($\sim$3087 stars) received a complete 
parametrisation in \T, \g, \meta\ and \AF. Radial velocities (\Vrad) were determined for 56\% of these FEROS  spectra. The non-parametrised spectra correspond to instrumental issues, 
overly low signal-to-noise ratios (S/N) and/or overly low level of quality overall for the analysis. Moreover, some
of these rejections concern hot and/or rapidly rotating stars incompatible with the reference grid adopted within AMBRE \citep{deLaverny2012}, which has been optimised for slow-rotating FGKM stars. Further details are available in \cite{worley2012}.

\section{Determination of the line-broadening velocities}
\label{Sec:Method}
\subsection{Adopted methodology}
For the \Vb\ determination within the AMBRE Project, a cross-correlation function is computed between the FEROS spectra, $S(\lambda)$, and a specific binary mask,  $M(\lambda)$. The binary masks consist of selected thin atomic lines and are generated from theoretical spectra representing the main expected absorption lines for a given stellar type. In the following, we adopted the masks specifically defined for AMBRE by \cite{worley2012}
for the determination of the radial velocities. More specifically, the spectrum emitted by a rotating star can be written as
\begin{equation}
S(\lambda) = H(\lambda) \ast G(\lambda),
\label{equ:stellar:rotation}
\end{equation}

where $H(\lambda$) is the spectrum emitted by a non-rotating stellar surface and $G(\lambda)$ is the rotational broadening function as described in \cite{Gray2008}, whose main parameter is the stellar rotation rate, \Vsini. Here, $G(\lambda)$ also varies with the limb-darkening coefficient: the \cite{Claret2000} values were adopted hereafter. The cross-correlation function with the adopted mask, $M(\lambda)$, is
\begin{equation}
CCF(\lambda) = \int_{0}^{\infty} \, S(\lambda) \cdot M(\lambda)\ d\lambda,
\label{equ:ccf}
\end{equation}
which can also be expressed as a function of the velocity, $v$, as $CCF(v)$.
For computational efficiency and following \citep{worley2012}, we restricted the integration interval to $\pm$500~\vunits. 
Two velocity steps ($\sim$2.22 or 4.44~\vunits) for the integration of Eq.~\ref{equ:ccf} were considered since two different setups are available for the FEROS spectra: either a wavelength sampling of 0.003 or 0.006~nm, respectively.

Subsequently, we determined the  $CCF(v)$ minimum location (i.e. equivalent to a \Vrad\ measurement), 
adopting a Gaussian fitting. More importantly, the full width at half maximum ($FWHM_{CCF}$ and associated uncertainty) of this Gaussian was derived. It is proportional to the \Vb\ that we want to determine, that is itself proportional to \Vsini\ \citep[see the introduction and][]{Recio2004, Melo2001, Weise2010}. These quantities are related via
\begin{equation}
    V_{\rm broad} = A \sqrt{\sigma_{CCF}^{2} - \sigma^{2}_{0}}
    \label{equ:vb}
,\end{equation}
where $A$ is a coupling constant that is specific to the considered spectral type (see Sect.~\ref{sec:coup:const}). Moreover, $\sigma_{CCF}$ is the width of the CCF,
computed from $\sigma_{CCF} = \frac{FWHM_{CCF}}{2\sqrt{2ln(2)}}$. Finally, $\sigma_{0}$ corresponds to
the CCF width for a non-rotating star and is therefore proportional to all the other broadening mechanisms, including the instrumental line spread function, except rotation.

\subsection{Determination of $\sigma_{0}$}\label{sec:sig0}
The CCF widths for a non-rotating star ($\sigma_{0}$) are specific to a given spectral type. They were derived  by solving Eq.~\ref{equ:vb} for synthetic spectra computed without considering any rotation and mimicking FEROS observed spectra in terms of resolution and sampling. Thus, $\sigma_{0}$  measures the mean spectral line broadening caused by intrinsic stellar effects (i.e. those inherent to the stellar atmosphere, independent of rotation) and instrumental effects. 

For that purpose, we adopted the AMBRE synthetic spectra \citep{deLaverny2012}, which were first generated at very high spectral resolution (R = 150\,000), without radial or rotational velocities. They covered the wavelength range between 300 - 1200 nm with a step of 0.001~nm. A sub-sample of this large grid was selected for the present study in order
to consider spectra representative of the AMBRE/FEROS parameters (FGKM-spectral types):
4000~K $\leq$\T$\leq$8000~K (step of 250~K); 1.0$\leq$\g$\leq$4.5 (step of 0.5~dex, $g$ in cm/$s{^2}$); -3.0$\leq$\meta$\leq$+1.0~dex (step of 0.5~dex); as detailed in Table 2 of \cite{deLaverny2012}. We also selected those spectra whose metallicity and $\alpha$-element  abundances satisfy the classical chemical pattern observed in the Milky Way: \AF=0.0~dex for~\meta$\geq$0.0~dex; \AF=-0.4$\times$\meta~for -1.0$\leq$\meta$\leq$0.0~dex, and \AF=+0.4~dex for \meta$\leq$-1.0 dex. 
We note that some combinations in the parameter space are not available, leading to the selection of 753 spectra (hereafter called AMBRE$_{Vbroad}$). 

To simulate ESO/FEROS spectrograph observations, these AMBRE$_{Vbroad}$ spectra were then subject to a two-step process. First, a Gaussian instrumental broadening was applied to match the FEROS resolution (R = 48\,000), followed by a resampling at the two FEROS wavelength steps (0.003~nm and 0.006~nm). Finally, the $\sigma_{0}$ were derived by computing the $CCF(v)$ 
between each AMBRE$_{Vbroad}$ spectra and each mask of \cite{worley2012}. 
We selected the optimal mask for each spectrum by only considering the closest mask and spectrum parameters.
We checked that the Gaussian fit of the 753 $CCF(v)$ is well centred 
at 0$\pm$2.5~\vunits\ since the spectra are at rest.
We also note that the selected $\sigma_{0}$ were always found to be larger than the FEROS instrumental broadening which is estimated to be close to 
$\sim$2.3~\vunits\ (corresponding to $R$=48\,000). The $\sigma_{0}$ uncertainties were directly estimated from the $FWHM_{CCF}$ ones. The median of these $\sigma_{0}$ relative uncertainties is 2.3\% with a dispersion of 1.9\%.

\subsection{Determination of the coupling constants, $A$}\label{sec:coup:const}
Once the $\sigma_{0}$ and their associated uncertainties are known, the determination of the coupling constant, $A$, first requires the computation of the $\sigma_{CCF}$. 
For that purpose, we applied a similar procedure as above, but
considering rotating AMBRE$_{Vbroad}$ spectra, adopting 18 \Vsini\ values (of 1, 2, 4, 5, 7, 10, 13, 16, 20, 25, 30, 35, 40, 50, 60, 70, 80, and 100~\vunits) for the $G(\lambda)$ broadening function. The simulated rotating spectra (see Eq.~\ref{equ:stellar:rotation}) were computed thanks to the \text{PyAstronomy rotBroad} function\footnote{Implemented in the SciPy library, see \url{https://pyastronomy.readthedocs.io/en/latest/pyaslDoc/aslDoc/rotBroad.html}}. We
 then simulated FEROS-like spectra as above by considering the instrumental profile and sampling.
The coupling constants were finally derived by fitting the (\Vb-$
\sigma_{CCF}$) relation of Eq.~\ref{equ:vb}, 
using the {\it scipy curve-fit} method\footnote{\url{https://docs.scipy.org/doc/scipy/reference/generated/scipy.optimize.curve_fit.html}}. For the uncertainty estimation, we conducted a series of 1000 Monte Carlo (MC) simulations considering both the $\sigma_{0}$ and $
\sigma_{CCF}$ uncertainties, deriving the $A$ value for each of them. We then adopted for $A$ the median of the distribution of all 1,000 computed coupling constants, along with its  uncertainty, which is half the difference between the 84$^{th}$ and 16$^{th}$ quantiles
of the distributions (thereby corresponding to a 1-$\sigma$ uncertainty for a Gaussian distribution).

\begin{figure}[t]
\centering
\includegraphics[width=.95\linewidth]{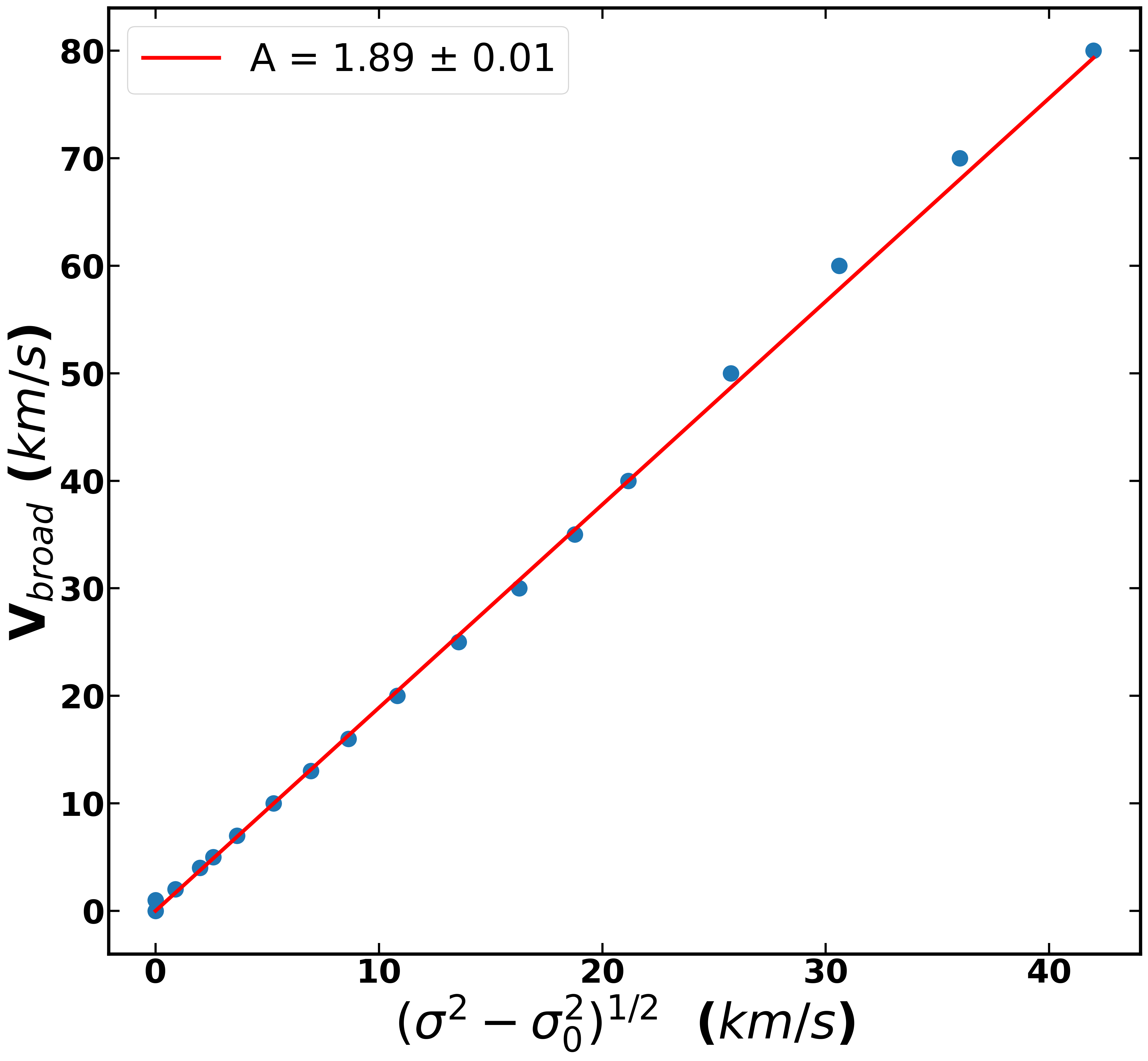}%
\caption{Coupling constant of Eq.~\ref{equ:vb} for the Solar spectrum and a mask having exactly the same parameters (FEROS-like spectrum with a wavelength sampling of 0.003 nm). The $A$-value and its uncertainty are reported in the upper-left corner.}
\label{Fig:slope_plots}
\end{figure}

We finally considered only the $A$ coupling constants obtained when $\sigma_{CCF}$ was estimated
for at least half of the \Vsini~values, ensuring a robust fit of Eq.~\ref{equ:vb}.
As above, we also retained only the cases for which the spectra parameters are the closest to the mask ones. This led to the derivation of 711 and 710 $A$-values for the 0.003~nm and 0.006~nm wavelength sampling, respectively. The median of their distribution is 2.02 (1.99) with a dispersion of 0.42 (0.44) for the 0.003~nm (0.006~nm) wavelength step. 
They are associated with relative uncertainties smaller than $\sim$3.12\% with  median values equal to 1.89\% and a dispersion of 2.59\%.
An illustrative example is shown in Fig.~\ref{Fig:slope_plots} for the Solar case (\T = 5750~K,  \g = 4.5, \meta = 0.0~dex, and \AF = 0.0~dex).

\begin{table*}[ht]
\caption{The AMBRE/FEROS catalogue of line-broadening velocities.}
\label{Tab:Vbroad}
        \centering
        \begin{tabular}{lcccccccccc}
                \hline 
                Gaia DR3 ID & \T & \g & \meta & \AF & \Vb & err\_\Vb & Upp\_Lim & \SNR & $\chi_{\rm AMBRE}^{2}$ & $HQ$\\
                 & (K) &  &  (dex) & (dex) & (\vunits) & (\vunits) & & & &\\
                 \hline
                 \Gaia\ DR3 2456361448068627712 & 5213 & 4.52 & 0.36 & -0.02 & 2.3 &  0.47 & N & 64.0 & -1.91 & Y\\
                \Gaia\ DR3 3031022299846117888 & 5422 & 4.24 & 0.14 & 0.16 & 5.41 &  0.48 & N & 70.0 & -1.93 & Y\\
                 \Gaia\ DR3 4623036865373793408 & 6013 & 4.43 & 0.1 & 0.04 & 3.11 &  & Y & 116.0 & -2.58 & Y\\
                 \Gaia\ DR3 4055808570473090560 & 4982 & 2.94 & -0.28 & 0.18 & 11.49 &  0.23 & N & 63.0 & -2.03 & Y\\
                \Gaia\ DR3 5835498883232034816 & 4973 & 4.56 & -0.44 & 0.17 & 3.06 &   & Y & 145.0 & -2.42 & Y\\
                ... & ... & ... & ... & ... & ... & ... & ... & ... & ... & ...\\
                \hline
        \end{tabular}
        \tablefoot{The full version of this table is available in electronic form at the CDS. The columns 'Upp\_Lim', '\SNR' and, '$\chi_{\rm AMBRE}^{2}$' refer to upper limit measurements (Y) for which no uncertainty are derived, the FEROS spectra S/N values (averaged in case of repeats) and the quality of the fit between the observed and reconstructed synthetic spectrum at the AMBRE parameters (also averaged when repeats), which is a good indicator of the parameterisation quality, respectively. The last column $HQ$ refers to the High-Quality measurements, as defined in Sect.~\ref{Sec:Discussion}.}
\end{table*}

\subsection{\Vb\ results for the AMBRE/FEROS spectra}
By applying Eq.~\ref{equ:vb}, the spectral line broadening was computed for the 
5,448 FEROS spectra of \cite{worley2012} having a published value for \T, \g, \meta\footnote{For the very few stars without an \AF\ estimate, we adopted the classical Galactic disc relation as in Sect.~\ref{sec:sig0}},
excluding the overly cool (\T $\la $3,500~K) or too hot ($\ga$8000~K) ones and those with the lowest quality parameter parameterisation (AMBRE/CHI2\_FLAG >1), since we could otherwise have ended up analysing them with a wrong mask. These spectra are associated to a $\sigma_{CCF}$ relative uncertainty  median value equal to 2.15\% with a dispersion of 2.2\%. For all of them, we adopted the most suitable mask (and therefore an $A$ constant)
by considering the closest mask and spectrum parameters,
as already performed in Sect.~\ref{sec:sig0}.
The uncertainties were again estimated thanks to a series of 1000 MC simulations, propagating the uncertainties on $A$, $\sigma_0$, and $\sigma_{CCF}$.
For several spectra, we found that $\sigma_{CCF}\simeq\sigma_0$ (within error bars), revealing an overly slow rotating star to be measured with the FEROS instrumental resolution. We therefore adopted for these spectra a
conservative \Vb\ upper limit, defined as \Vb <  A$\sqrt{2\sigma_{0} \times\epsilon_{CCF} +  \epsilon_{CCF}^2 }$, where $\epsilon_{CCF}$ is the error associated to $\sigma_{CCF}$, and checking to ensure that this value is always larger than the FEROS instrumental broadening. 
This procedure led to the determination of 2349 \Vb\ values and 3099 upper limits.
The median of the relative uncertainties (excluding the upper limits for which no uncertainties are estimated) is 8.1\%,  with a dispersion of 9.8\%. 

\section{The AMBRE/FEROS catalogue of line-broadening velocities}
\label{Sec:Catalog}

The sample of 5,448 spectra with a derived line-broadening velocity actually corresponds to a smaller number of stars, since several of them have been observed at different epochs. To build the AMBRE/FEROS catalogue of \Vb, the first step was to carefully identify the star corresponding to each spectrum. For that purpose, we followed a similar
procedure as in \cite{Santos-Peral2021} and identified the \Gaia\ Data Release 3 (DR3) ID number for each source. 
We recovered the \Gaia\ IDs of 5,331 spectra corresponding to 2,580 different stars. We also retained 117 other spectra 
that are associated with four other stars being too bright for having a \Gaia\ ID and for which we adopted their Henry Draper (HD) identification.
For the few remaining cases, we were not able to securely
identify their corresponding star because of the rather low accuracy of the FEROS spectra coordinates. These were disregarded hereafter. 

We recall that all these spectra have high-quality AMBRE atmospheric parameters as their parameterisations are associated to
quality flag AMBRE/CHI2\_FLAG values equal to 0 or 1.
First, among them, 1957 stars have only one FEROS spectrum, and thus a unique computed \Vb\ and associated uncertainty values. 
Secondly, 627 other stars have more than one derived \Vb\ value because several repeat\footnote{We recall that the AMBRE Project parameterised individual archive spectra and, for some stars, several spectra could have been collected, leading to several parameter sets.} FEROS spectra were analysed for them. 
When two repeat spectra are available, we simply averaged the two derived \Vb\ 
and uncertainty values, after having checked that the spectra S/N and parameterisation were close within each other. When more than three spectra were analysed for a given star, we applied a distance-weighted averaging method as in \cite{Adibekyan2015, Perdigon2021}, assigning a larger weight to spectra with higher S/N. A similar procedure was adopted to compute the mean \Vb\ uncertainties, the mean atmospheric parameters and the mean quality of the fit between the observed and the reconstructed synthetic spectrum at the AMBRE parameters \citep[$\chi_{\rm AMBRE}^{2}$, see][]{worley2012} for these stars with several repeat spectra. 

All of this led to the \Vb~derivation for 2,584 AMBRE/FEROS stars. Their AMBRE atmospheric parameters \citep[as provided by][or averaged with weighs in case of repeat spectra]{worley2012} together with their line-broadening velocity and associated uncertainty are provided in Table~\ref{Tab:Vbroad}.  
We recall that no 
relative uncertainties are provided for upper limit values and, for these 1475 stars, the FEROS spectra S/N ratios could be considered for selecting the best quality estimates. 
We also recommend to adopt the fit quality ($\chi_{\rm AMBRE}^{2}$) to select the best data.
Finally, 75\% of stars with available uncertainty are associated to a relative error less than 10\% (96\% having a relative error less than 30\%). Moreover, we warn that our \Vb\ values are derived from the AMBRE atmospheric parameters (via the mask choice), which are assumed to be exact for the present study.

\section{Comparison with the literature}
\label{Sec:Valid}
The AMBRE/FEROS catalogue of line-broadening velocities is validated 
thanks to literature values, considering five main \Vb\ catalogues. The cross-match between these catalogues and ours was made
by adopting the \Gaia\ DR3 or the HD (four stars) identifications. These comparisons are presented by decreasing number of stars in common between the AMBRE and literature catalogues.
\subsection{\Gaia\ DR3 line-broadening parameters}
\begin{figure}[t]
    \centering
    \includegraphics[width=0.95\linewidth]{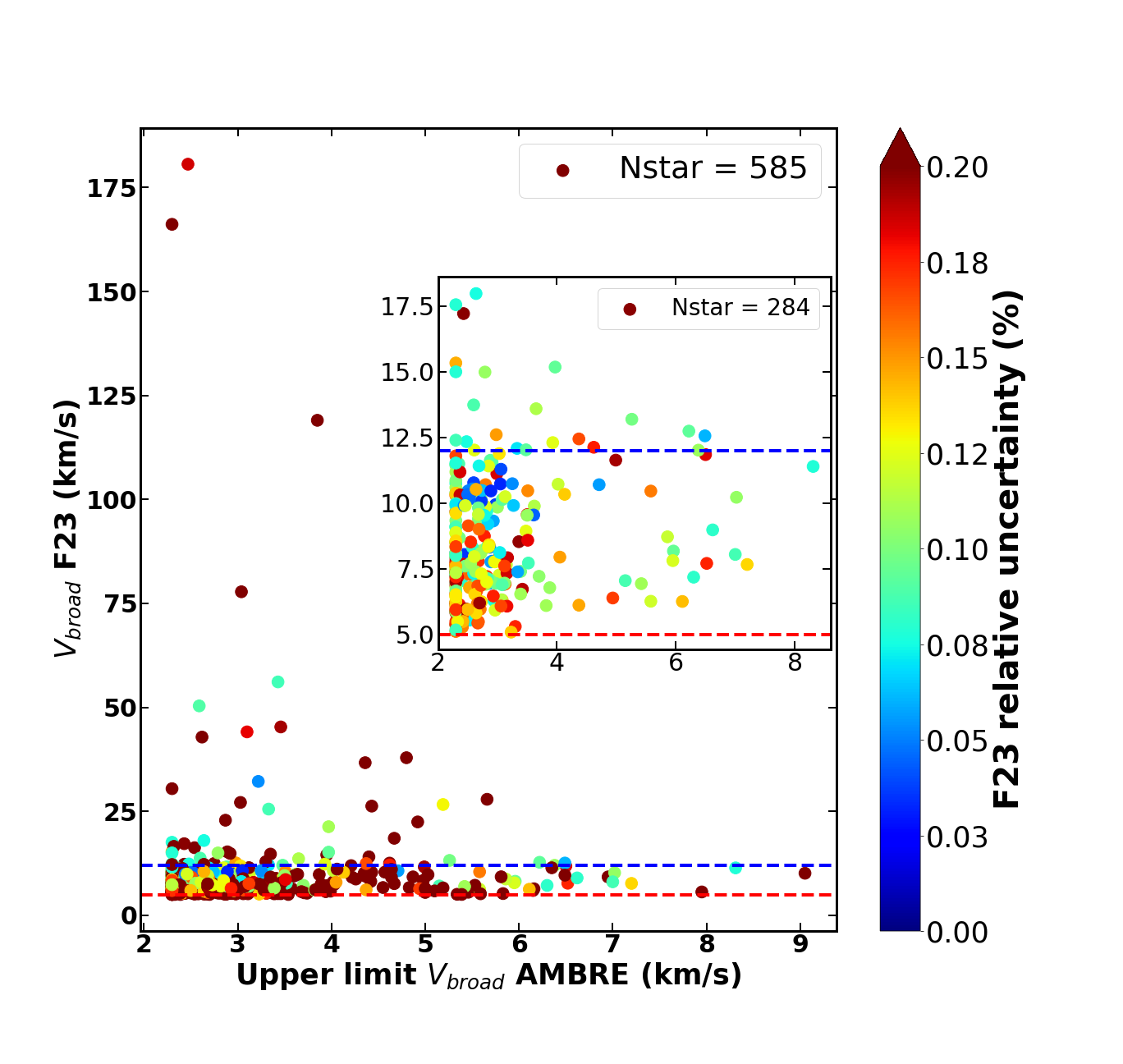}     
    \caption{Comparison of the AMBRE/FEROS \Vb\ upper limits with the measurements of \cite{Fremat2023}, derived from \Gaia/RVS DR3 spectra. The horizontal red and blue  dashed lines refer to the F23 measurement limit and to their low-\Vb\ domain, where overestimated values are suspected (5 and 12~km/s, respectively). The colour coding represents the F23 \Vb\ relative uncertainty. The inset focuses on low-rotating F23 stars having a \Vb\ relative uncertainty lower than 20\%.}
    \label{Fig:ValidGaia}
\end{figure}
\begin{figure}[ht]
    \centering
    \includegraphics[width=0.95\linewidth]{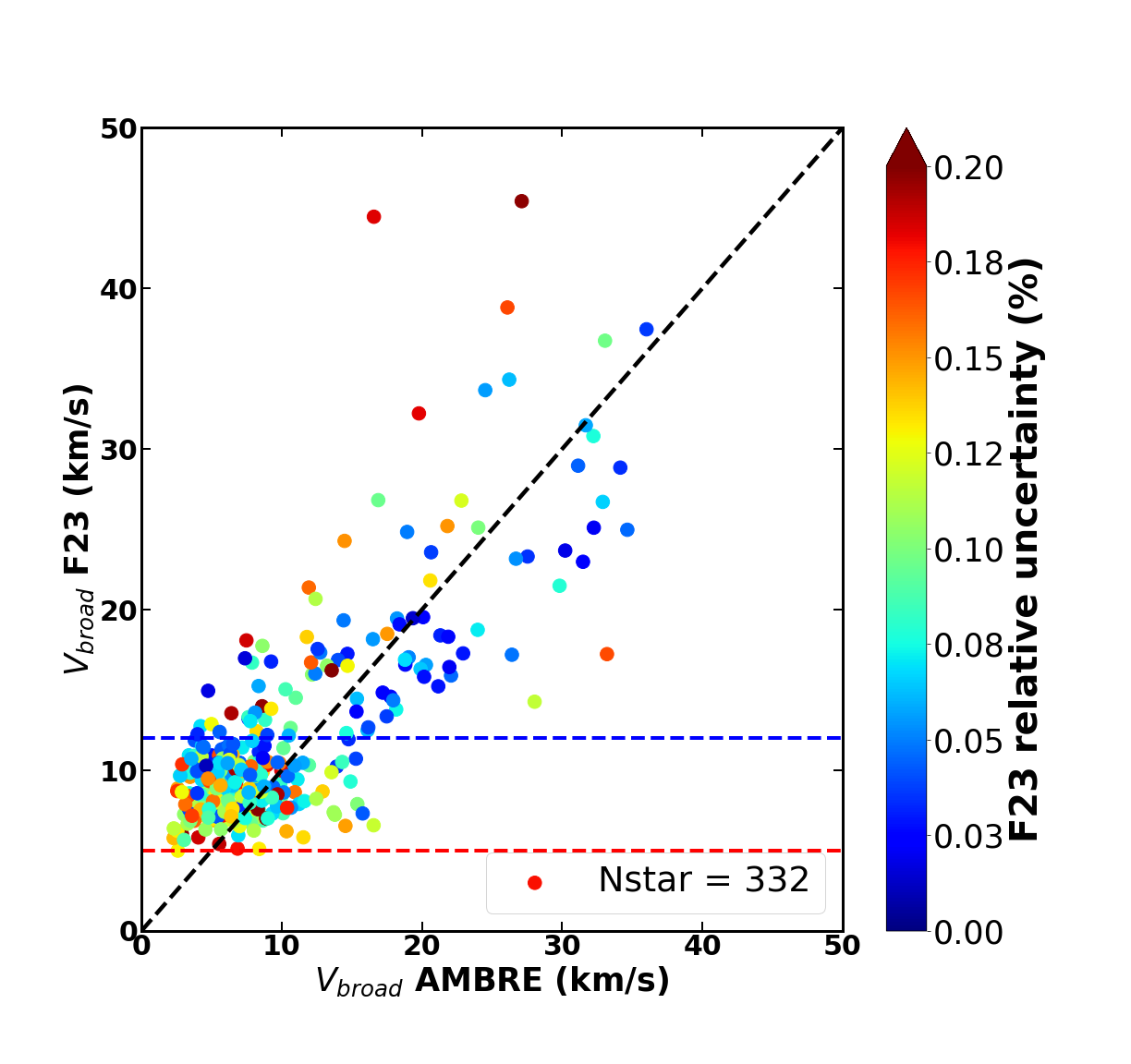}     
    \caption{Same as Fig.~\ref{Fig:ValidGaia} but considering only AMBRE stars with a fully derived \Vb\ (i.e. excluding upper limits) and high-quality measurements in both catalogues (\Vb\ relative uncertainty less than 20\%). 
    The black-dashed line denotes the one-to-one relation for visual reference. For clarity, six extreme F23 stars with \Vb>50~km/s are not shown (see text for details, including a discussion about the bias and dispersion associated to this comparison).}
    \label{Fig:ValidGaia2}
\end{figure}

\citet[][F23, hereafter]{Fremat2023} measured the line broadening parameter (i.e. equivalent to our \Vb) of about 3.5 million stars by analysing their \Gaia's Radial Velocity Spectrometer DR3 data \citep{RVS18}. Their huge catalogue 
is therefore based on spectra around the Ca~II IR triplet lines. These spectra resolution ($R\sim$11,500) is much lower than the FEROS one and their sample is limited to G$_{RVS}<$12. We found 
1172 stars in common between F23 and our catalogue and we note that our relative uncertainties are always smaller than the F23 ones.  For the stars in common, the FEROS (RVS) S/N have a mean equal to 95 (361) and 1162 (1130) stars are characterised by S/N>50. Only two stars have a FEROS S/N<40, but rejecting them did not change our conclusions, presented hereafter.

We first show in Fig.~\ref{Fig:ValidGaia} a comparison between our upper-limit \Vb\ values and the F23 measurements. For these 585 stars, it can be seen that most F23 stars are confirmed as slow rotators, in particular those with rather low relative uncertainties (inset). Our upper limits value are however always lower than the F23 values. This confirms the F23's claim that their smallest values (below $\sim$12~km/s, blue dashed line in Fig.~\ref{Fig:ValidGaia}) are overestimated partly because of the rather low RVS spectral resolution. There are however some F23 stars with very high \Vb\ values that are not compatible with our much lower upper limits.
The most extreme cases (with \Vb(F23)>70~\vunits\ and all associated to rather F23 large relative uncertainties) are 
\Gaia~3030262502952031616, 5201479762265799168, 
3877936489933697408, and 5732039576302762496 having \Vb(F23) equal to 181, 166, 119, and 78 respectively, whereas our upper limits are lower than $\sim$4 \vunits . 
For all these stars, we first checked that their FEROS and RVS spectra S/N are of rather high quality and that their AMBRE parameterisation is in agreement with literature values (when available), confirming
the good choice determined when selecting the masks for Eq.  \ref{equ:ccf}. 
No peculiar information (spectral type and rotation) was found in the literature for the two first stars, although the first one could
be a suspected binary belonging to the NGC~2423 open cluster. The third one is a known double or multiple star, according to CDS/Simbad\footnote{https://simbad.u-strasbg.fr/simbad/}, for which \Gaia\ identified only one source. Hence, its RVS spectrum could be affected by different stellar contributions leading to a possible suspicious \Vb\ determination.
Finally, \Gaia~5732039576302762496 is a short period pulsating star,  bright in X-rays, and, thus, with a possible complex spectrum that could have been poorly analysed. 

Our \Vb\ measurements without considering the upper limits are compared to F23 in Fig~\ref{Fig:ValidGaia2}. For clarity reasons, we only show in this figure all the high-quality measurements defined as having a \Vb\ relative uncertainty less than 20\% in both catalogues (in practice, this actually filters out almost exclusively F23 stars since our uncertainties are smaller). 
We note, however, that four stars are not shown in Fig.~\ref{Fig:ValidGaia2} because their \Vb(F23)$\gg$50 \vunits. These extreme stars are commented hereafter and are not considered in the statistical comparison below.
The global agreement looks good, in particular if we consider that (i) the F23 measurements below $\sim$12 \vunits\ are probably overestimated (see discussion above) and (ii) the largest disagreements are found for the largest F23 uncertainties. The mean difference between both catalogues, considering the 332 stars with relative uncertainties less than 20\% is indeed 3.8~km/s with a dispersion of 3.0~\vunits. 
This corresponds to a mean relative difference of 33\% and
a relative dispersion of 22\%.
However, these numbers quickly increase when considering F23 stars with larger uncertainties:
there are indeed 25 stars with \Vb(F23)$\ga$50~km/s and relative uncertainties  larger than 20\% (with nine having an uncertainty larger than 40\%).

Among the four stars not shown in Fig.~\ref{Fig:ValidGaia2} because of
their large \Vb(F23) and with the other dozen having F23 relative uncertainties smaller than 40\% (but departing by more than 40~\vunits\ with respect to our determination), some suspicious values are present in one of the catalogues. Some of these cases could come
from a wrong AMBRE parameterisation that led to a wrong choice of the mask for Eq.~\ref{equ:ccf}, but others could also be low-quality F23 determination. As example, we have identified among the most discrepant cases, \Gaia~4946938113149426944 with \Vb(F23)=49.5$\pm$10.7~\vunits, whereas we have derived \Vb=3.3$\pm1.1$~\vunits.
The AMBRE stellar atmospheric parameterisation of this star fully agrees with previous determination \citep[see, e.g.][]{Costa21}. Moreover, literature \Vsini\ values are found around $\sim$15~\vunits, which strongly depart from F23, despite the fact they are still larger than ours. Actually, this star is a well known chromospheric active star. Since the RVS spectrum analysed by F23 is strongly dominated by the huge Ca~II triplet lines, it is highly possible that the F23 measurement could have been affected by chromospheric effects in these line profiles. Finally, when examining its FEROS spectrum, a large \Vb\ is excluded and we are therefore confident with our lower determination that relies much less on lines affected by the chromosphere thanks to the considered mask and the wider spectral domain studied. Moreover,
the examination of the FEROS spectrum of \Gaia~6054867464551287168 could reveal a binary component, affecting both the AMBRE and F23 determinations
(5.3 and 234.6~\vunits, respectively). The FEROS spectrum of \Gaia~5935941572487329536 also exhibits some line core emission that could indicate some chromospheric activity,
affecting more the F23 analysis. For this star, we however derived \Vb=16.6~\vunits\ in good agreement with \cite[][20~\vunits]{Torres06}, contrary to \Vb(F23)=44.5~\vunits. 
Similar chromospheric activities associated to very large (and, thus, suspicious) \Vb(F23) values are also seen in the spectrum of 
\Gaia~4153657378750644864, \Gaia~4114430411648869632 \citep[our value being closer to the one of][]{Torres06}, and \Gaia~5369414461016305408 \citep[our value being in good agreement with the one of][]{Torres06}. 

\subsection{Glebocki \& Gnacinski's compilation of rotational velocities}
\begin{figure}[t]
    \centering
    \includegraphics[width=0.95\linewidth]{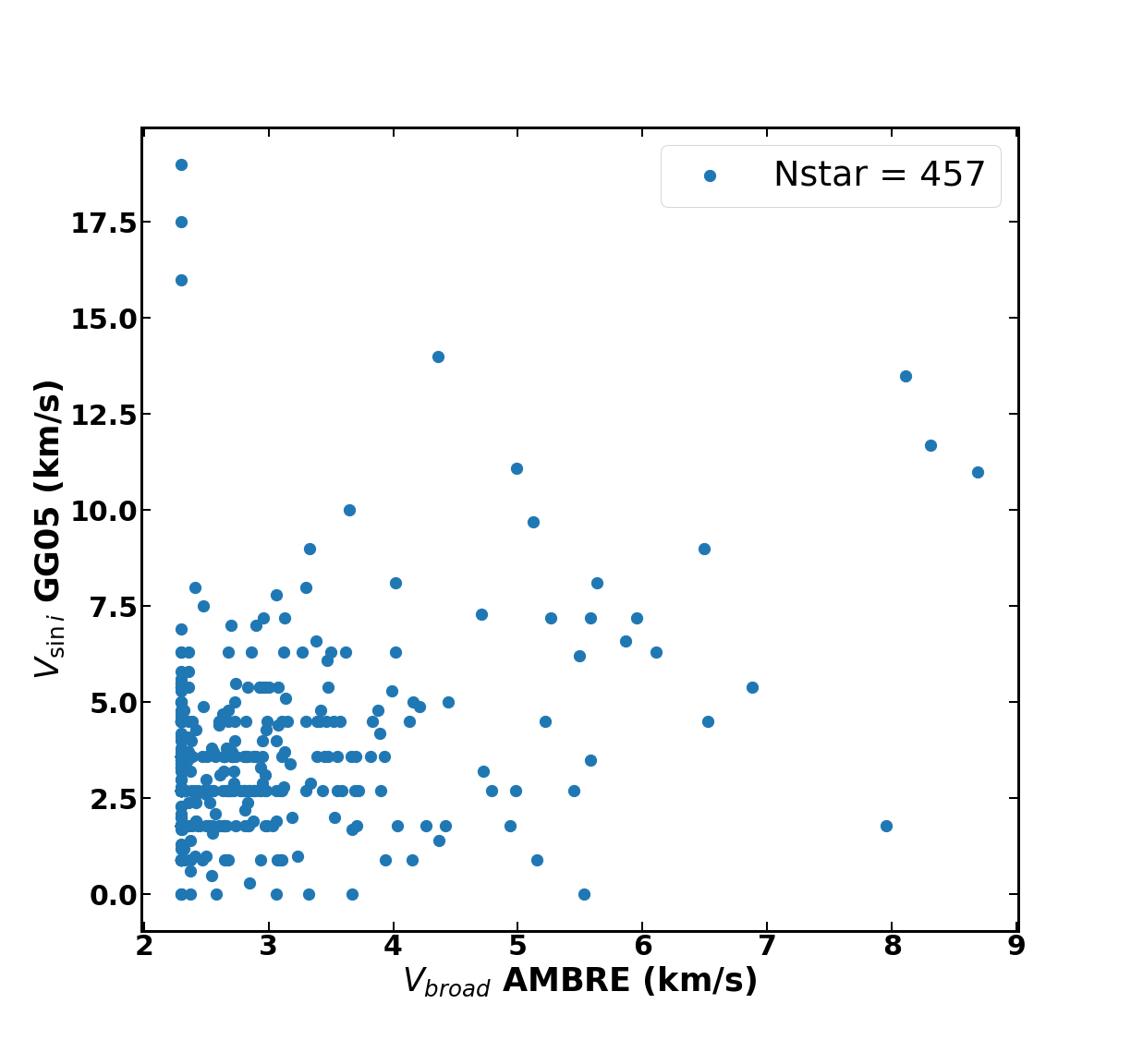}     
    \caption{Comparison of the AMBRE/FEROS upper limits \Vb\ with the measurements of \cite{GG2005}.}
    \label{Fig:ValidGG05}
\end{figure}
\begin{figure}[t]
    \centering
    \includegraphics[width=0.95\linewidth]{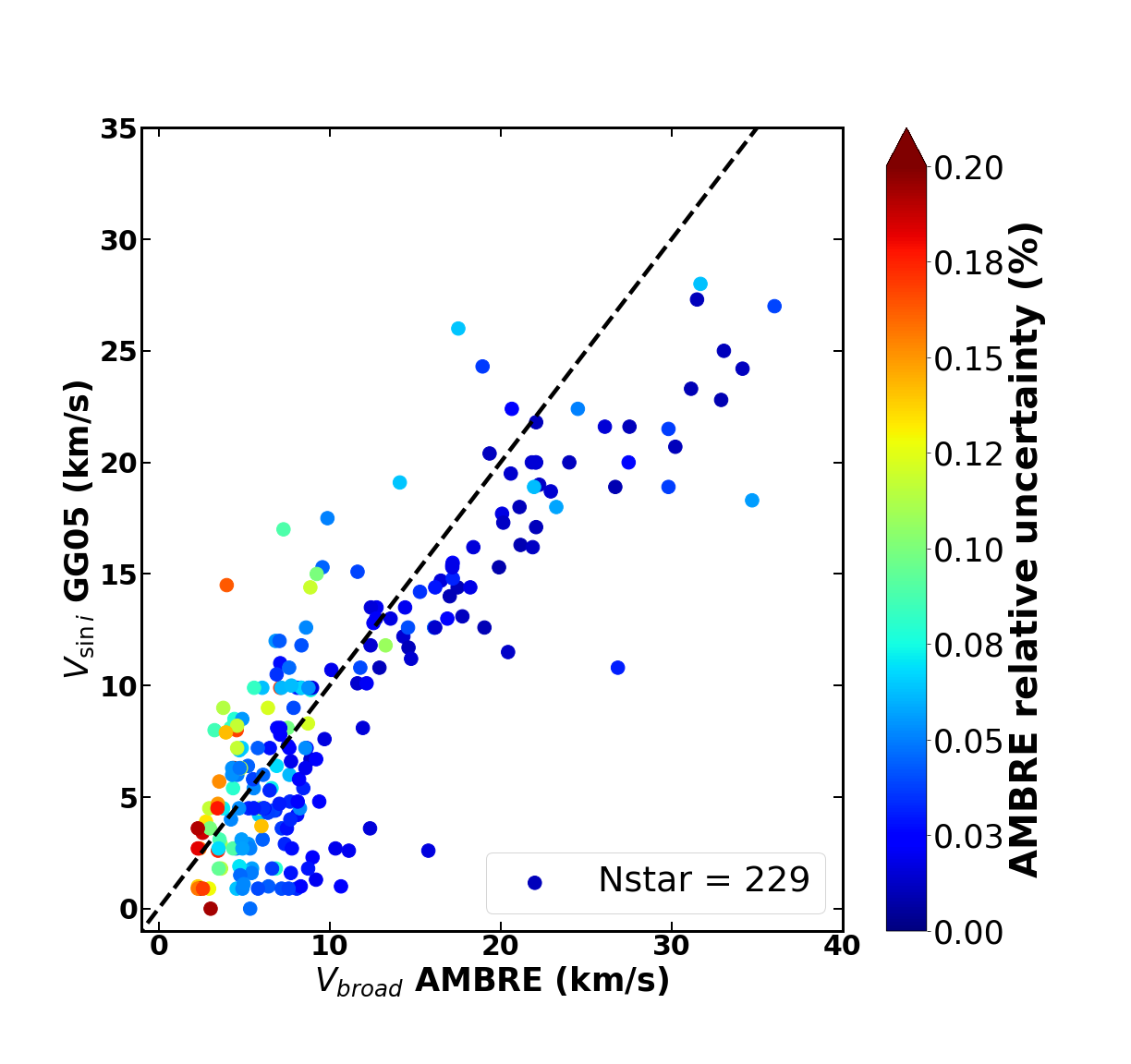}     
    \caption{Same as Fig.~\ref{Fig:ValidGG05} but excluding the AMBRE/FEROS upper limits, and only considering AMBRE/FEROS high-quality measurements (relative uncertainties less than 20\%; see the colour coding). The black-dashed line denotes the one-to-one relation for visual reference. See the text for a discussion about the bias and
dispersion associated to this comparison.}
    \label{Fig:ValidGG052}
\end{figure}
The catalogue published by \citet[GG05, hereafter]{GG2005} is a compilation of projected rotational velocities (\Vsini\footnote{These values, if really corresponding to \Vsini\ measurements (i.e. without the contribution of the macro-turbulence, for instance) for every star, could be lower than our \Vb\ that also include other broadening mechanisms.}) published by various authors and collected between 1981 and 2000. The rotational information, that could be of rather heterogenous quality, are available for 28179 stars belonging mainly to 166 open clusters, globular clusters and associations. For each star, this catalogue provides the 
mean \Vsini\ value when several literature values are available.
We found 746 stars in common between the GG05 and AMBRE/FEROS catalogues. For the comparison, we only considered their stars having a good quality flag (empty {\it n\_vsini} flag), decreasing
the comparison sample to 722 stars. Among them, 256 stars have a \Vb(AMBRE) value, whereas 466 others have an upper limit estimation in the AMBRE/FEROS catalogue.

We first show in Fig.~\ref{Fig:ValidGG05} a comparison of all our \Vb\ upper limits with the GG05 values. 
Our estimates well confirm that most of these stars are slow rotators (below $\sim$10\vunits), as identified by GG05. We note, however, that no errors are provided by GG05 that could help to better understand the very few
discrepant cases. For instance, two stars are not shown in this plot since their \Vsini(GG05) is larger than 20~\vunits, whereas our upper limits are lower than a few \vunits. One of them, 
\Gaia~DR3 5201129121134048768 (\Vsini(GG05)=29.9~\vunits) is a known young star whose AMBRE parameters seem reasonable with respect to literature values. However, its complex spectrum that exhibits emission lines could have blurred one of the \Vb\ determinations (or both). On the other hand, GG05 report for \Gaia~DR3 6207412157262342784 \Vsini\ values between 19$\pm$6 and 38~\vunits. According to CDS/Simbad, this is a chemically peculiar star with, again, a rather good AMBRE parametrisation. However, their lowest value, even considering its uncertainty, is  marginally consistent with our upper limit. 

Figure~\ref{Fig:ValidGG052} illustrates the comparison with GG05 for stars with a fully derived AMBRE \Vb\ (i.e. no upper values are included in this figure), and a \Vb(AMBRE) relative uncertainty lower than 20\%. A good agreement can be seen with a mean difference between both catalogues equal to 3.3~\vunits\  and a dispersion close to 2.8~\vunits. 
It can however be seen in this figure that the GG05 values seem to be quite often slightly smaller than ours. This could be an effect of the different definition of \Vb\ and \Vsini: the macroturbulence velocity is indeed included in our estimates, contrarily to the GG05 ones that are believed (assumed)
to be \Vsini\ values, without considering any other broadening mechanisms.
Moreover, we note that two stars are not shown in Fig.~\ref{Fig:ValidGG052} because of their large \Vsini(GG05). 
Among them, \Gaia\  5194203331750283392 has \Vb(AMBRE)= 34.7~km/s, whereas
GG05 report a value $\sim$10~km/s larger. We note that its AMBRE stellar parameters are consistent with literature values found in CDS/Simbad. Additionally, \cite{Zuniga21} published a rather low \Vsini, closer to ours (28~km/s), which could, thus, be favoured. Then, \Vb(AMBRE)=26.3~km/s and \Vsini(GG05)=41.4~km/s for \Gaia\ 5199981334076455680, whereas
\cite{Zuniga21} reported 34.1~km/s. The stellar parameters of this star should be explored (i.e. confirmation of the AMBRE values) to better define its actual rotational rate.

\subsection{GALAH/DR4 rotational broadening}

\begin{figure}[t]
    \centering
    \includegraphics[width=0.95\linewidth]{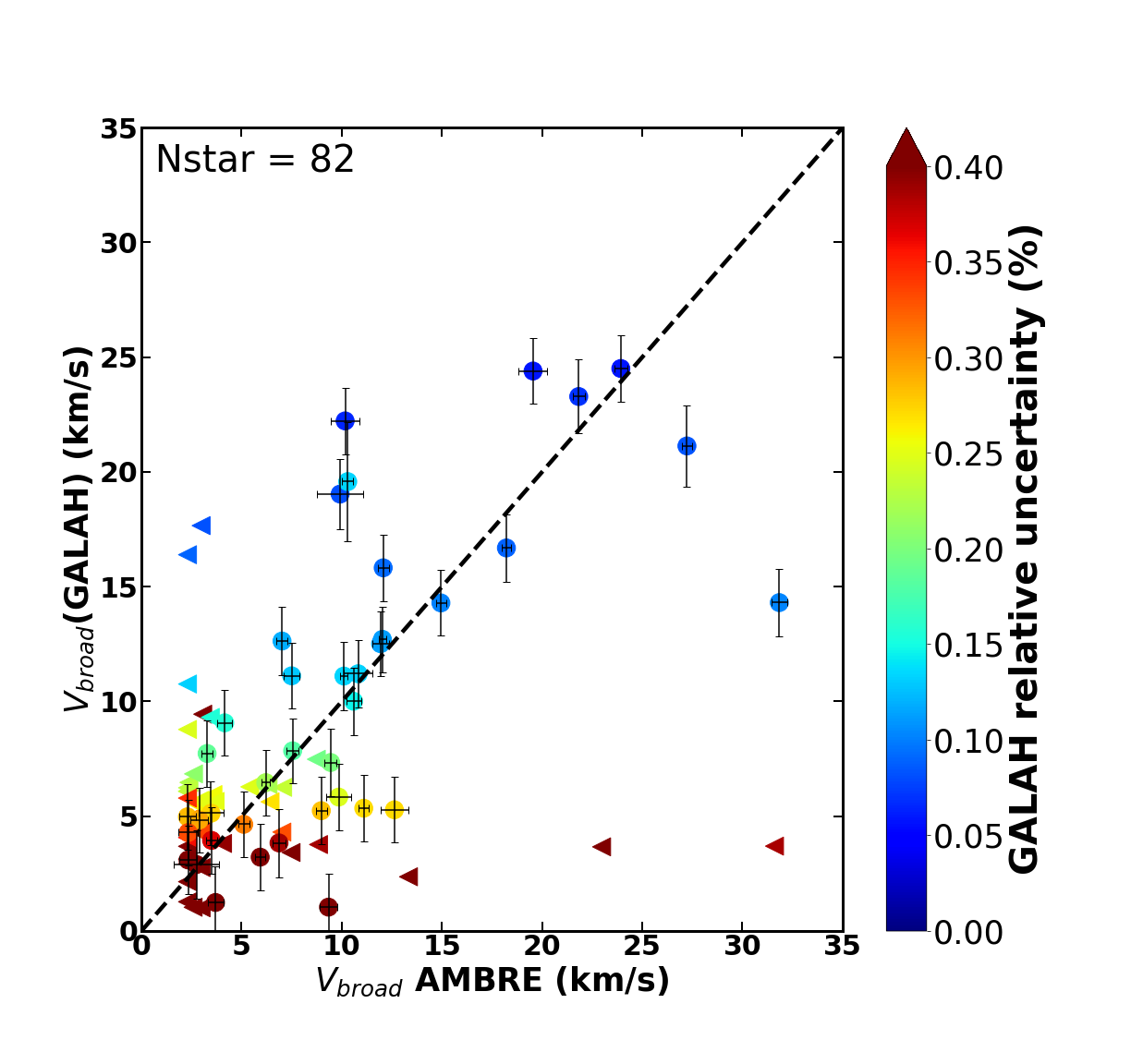}  
    \caption{Comparison of the rotational broadening AMBRE/FEROS \Vb\ with the GALAH measurements of \cite{Buder2024}. The black-dashed line denotes the one-to-one relation for visual reference. The colour coding represents the \Vb(GALAH) uncertainty. Left-pointing triangles refer to the AMBRE upper limits. See text for a
discussion about the bias and dispersion associated to this comparison.}
    \label{Fig:ValidG14}
\end{figure}

The GALAH project has published with its fourth data release a catalogue
of about 918,000 stars with a rotational information \cite[]{Buder2024}. For the analysis of the observed spectra ($R$=28,000), 
their reference synthetic spectra are computed without any rotational
and macro-turbulent broadening, and were then broadened by different \Vsini~values
to estimate \Vb. 

In the absence of specific quality criteria for their \Vb\ values, we adopted their quality flag (flag\_sp) associated to the spectroscopic parameterisation. Only common targets with flag\_sp = 0 were thus retained in our analysis.
We found 123 common stars; and 82 of them have flag\_sp = 0 and are considered hereafter. Among them, we have 43 AMBRE upper limits. We illustrate in Fig.~\ref{Fig:ValidG14} the comparison between their and our values for these 82 common stars. The agreement is rather good with a mean difference between both catalogues equal to 3.6~\vunits\  and a dispersion of 3.7~\vunits\ (not considering the 43 upper limits). Only two stars differ by more than $\sim$10~\vunits. More precisely, the difference in \Vb\ for \Gaia\ DR3 6242687277719794176 and 5524839912881567232 is 12.1 and 17.5~\vunits, respectively. The first one is a known K-type variable for which APOGEE published a \Vb\ value between that of AMBRE and GALAH \citep{Abdurrouf2022}. For the second one, the AMBRE and GALAH \T\ differ by about 900~K and the \Vb\ difference could then be explained by a misclassification in one of the two studies.

\subsection{APOGEE/DR17 rotational broadening}

\begin{figure}[t]
    \centering
    \includegraphics[width=0.95\linewidth]{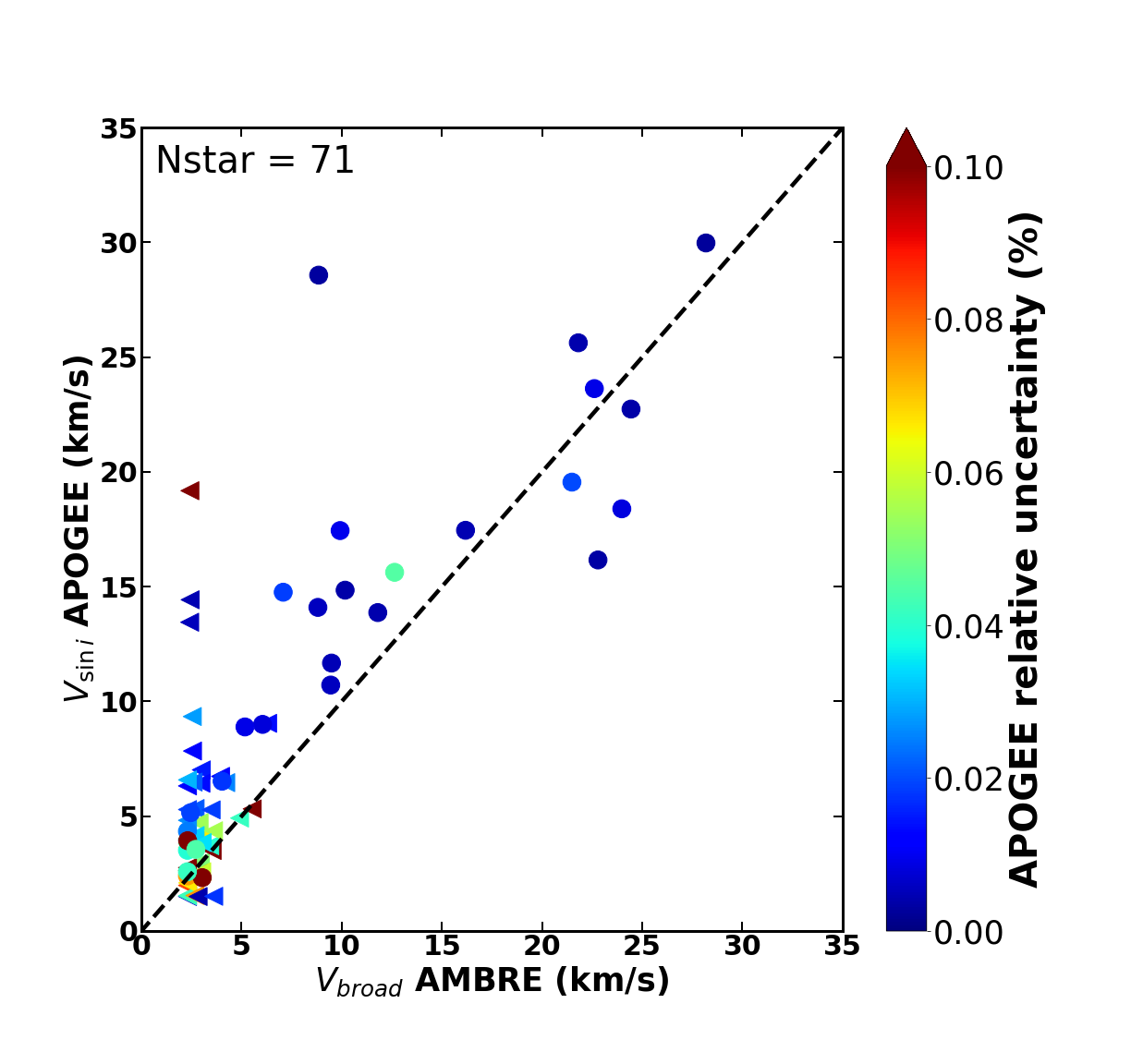}     
    \caption{Comparison of the rotational broadening AMBRE/FEROS \Vb\ with the DR17 APOGEE measurements of \cite{Abdurrouf2022}. The black-dashed line denotes the one-to-one relation for visual reference. Left-pointing triangles refer to the AMBRE upper limits. The colour coding refers 
    to the APOGEE relative uncertainties but most stars have overly small APOGEE uncertainties to see their error bars. See the text for a
discussion about the bias and dispersion associated to this comparison.}
    \label{Fig:ValidAPO17}
\end{figure}

The DR17 Apache Point Observatory Galactic Evolution Experiment (APOGEE)
contains the rotational velocity of about 
315,000 dwarf stars \cite[]{Abdurrouf2022}, derived from spectra having a resolution of 22,500. Since their adopted methodology derived the combined rotational and macro-turbulent velocities, they can be directly compared to our \Vb.

We identified 125 common stars between the APOGEE and AMBRE catalogues, but only 71 have an APOGEE rotational information associated to a reliable value, according to the APOGEE VSINI\_WARN, ROTATION\_WARN, and VSINI\_BAD quality criteria.
The remaining stars are shown in Fig.~\ref{Fig:ValidAPO17}, with 46 of them having an AMBRE upper limit (triangles in the figure). 
Except for one star (discussed below), the agreement without considering AMBRE upper limit stars is excellent with a mean difference of 3.4~\vunits, associated to a dispersion of 3.8~\vunits.
The star with the largest difference (about 20~km/s with \Vb(APOGEE)=28.6~\vunits) between APOGEE and AMBRE is \Gaia\ 3216112884666365440. It 
is a known spectroscopic binary according to SIMBAD. Its APOGEE and AMBRE \T\ and \g\ differ by about 500~K
and 1~dex, respectively. It could therefore be that there is a misclassification in one of the catalogue that could explain the \Vb\ disagreement.

The AMBRE upper limits are also in good agreement with the APOGEE values, except for four stars having a \Vb(APOGEE)\ga10\vunits\ (among 42).
One of them (\Gaia\ 4126229541458375808 with \Vb(APOGEE)=19.2~\vunits) has a huge uncertainty (16.1~\vunits) and could thus be in agreement with our upper limit value. The other three stars are: \Gaia\ DR3 3265655023187142784,  3827985985921614080, and 3374633977170787072 with \Vb(APOGEE)=13.5$\pm$0.07, 14.4$\pm$0.06~\vunits, and 9.4$\pm$0.26~\vunits~ respectively, whereas we derived \Vb\la2.5~\vunits\ for all of them.
The first one has a AMBRE \T\ hotter by about 400~K with respect to the APOGEE one, although other literature \T\ are very close to ours, hence favouring our derived \Vb. For the second one,
the \T\ differ by about 250~K but this would not explain the \Vb\ difference, which origin is unknown.
Finally, for the last star, we favour our low value since it was  confirmed by \cite{GG2005}.

\subsection{Rotational broadening of evolved stars from de Medeiros}

\begin{figure}[t]
    \centering
    \includegraphics[width=0.95\linewidth]{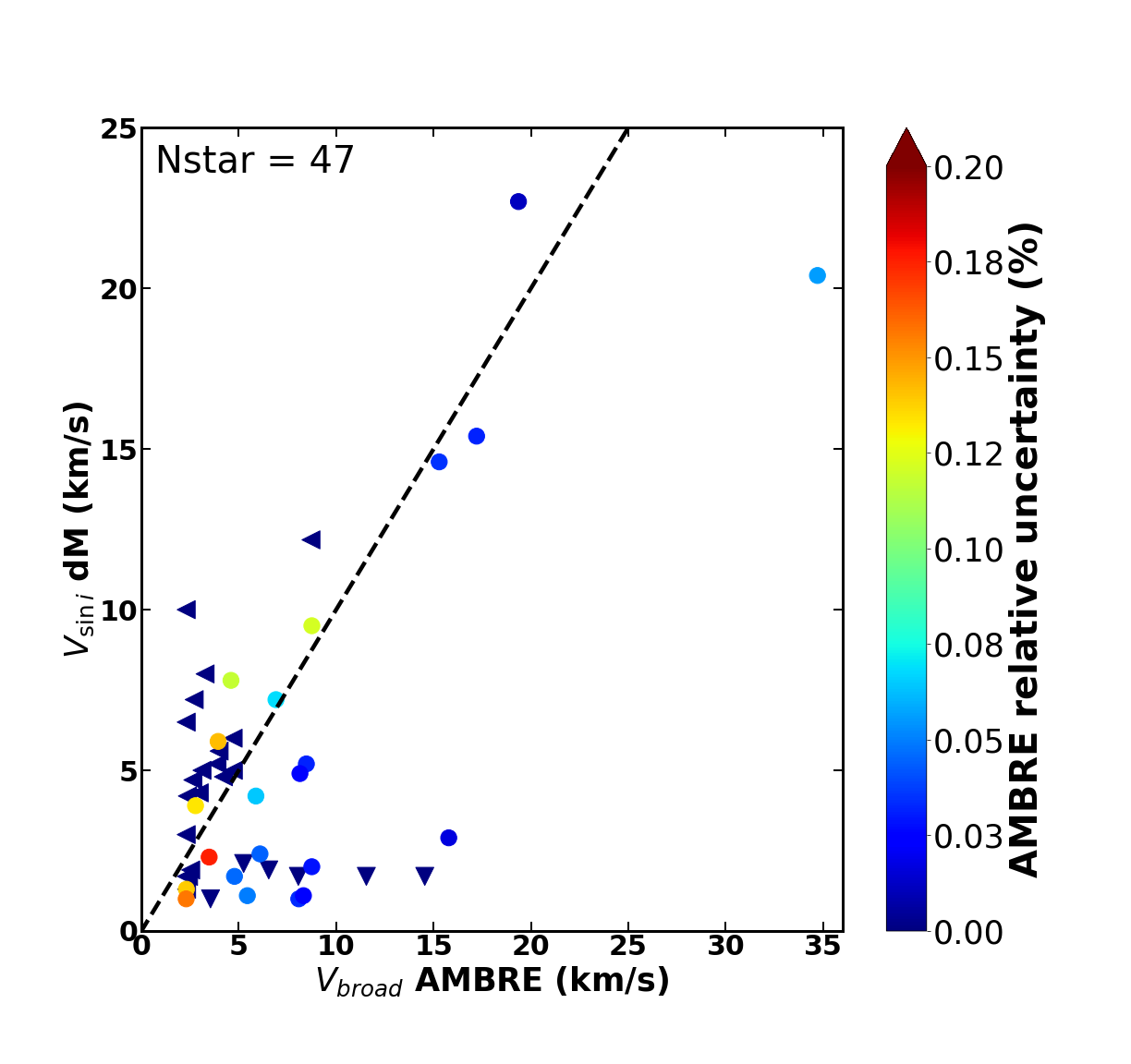}     
    \caption{Comparison of the rotational broadening of AMBRE/FEROS with the de Medeiros measurements. The black-dashed line denotes the one-to-one relation line for visual reference. The colour coding refers 
    to the AMBRE relative uncertainties. Left and downward pointing triangles refer to the AMBRE and dM upper limits, respectively. See text for a
discussion about the bias and dispersion associated to this comparison.}
    \label{Fig:ValidJRM}
\end{figure}

\begin{table}[t]
\caption{Number of the AMBRE/FEROS \Vb\ stars also found in the de Medeiros catalogues (no common stars were found with dM04).}
\label{Tab:Vbroad_JRM}
        \centering
        \begin{tabular}{cccc}
                \hline  \hline
                 de Medeiros article & Nstar & Upp\_Lim$_{\rm dM}$ & Upp\_Lim$_{\rm AMBRE}$ \\
                 \hline
                 dM99 & 35 & 03 & 11 \\
                 dM02 & 05 & 03 & 01 \\
                 dM06 & 07 & 00 & 07 \\
                  \hline
        \end{tabular}
\end{table}

The study of stellar rotation in evolved stars has strongly benefited from the pioneering works of~\citet[dM99, dM02, dM04, and dM06, hereafter]{JRM1999, JRM2002, JRM2004, JRM2006}. This series of papers provides databases for rotational velocities of evolved FGK stars. 
The observations were mostly collected with the CORAVEL spectrometer ~\cite[R=20,000,][]{Baranne1979}, enabling the measurement of \Vsini\  for a few thousands stars spanning luminosity classes Ib to IV.

We summarise in Table~\ref{Tab:Vbroad_JRM} the number of common stars between our catalogue and the de Medeiros ones. The rotational properties of the 47 common stars are compared in Fig.~\ref{Fig:ValidJRM}.
The agreement between both measurements is very good with 
a mean difference of 3.83~\vunits, associated with a dispersion of
3.68~\vunits\ (not considering the upper limits). The most discrepant star is \Gaia\ DR3 4156512638614879104 
with \Vsini(dM)=20.4~\vunits\ and \Vb(AMBRE)=34.71~\vunits, both having similar uncertainties close to $\sim$2~\vunits. According to CDS/Simbad and \Gaia, this is a single Cepheid variable.
Its AMBRE atmospheric parameters are very close to those 
reported by \cite{Caroline22} with a difference of $\sim$150~K , 0.5~dex, and 0.3~dex in \T, \g, and \meta\ respectively; this is associated with a rather good $\chi_{\rm AMBRE}^{2}$ quality parameter for a \SNR=75 spectrum. We are thus confident in our \Vb\ determination.

\subsection{Summary}
This section presents our validation of our \Vb\ determinations by comparing them with about 1,800 other literature measurements.
In particular, by comparing the AMBRE and \Gaia/DR3 catalogues: (i) we confirm that most of the F23 \Vb\ below $\sim$12~km/s are overestimated, (ii) both data sets agree within $\sim$4~km/s when considering F23 \Vb\ with rather low relative uncertainties (more than 300 stars being considered), and (iii) some F23 stars with \Vb$\ga$40~km/s are associated to rather large uncertainties that could be caused by chromospheric or binarity effects, preventing a proper derivation of \Vb. 
The agreement with the GG05, APOGEE, GALAH, and the de Medeiros catalogues  is also found to be good with again a mean difference of about 3.5~\vunits. 

Therefore, we concluded that the AMBRE/FEROS catalogue of rotational broadening is of high quality. 
 We recommend to well consider the relative uncertainties and the different associated quality flags to define high-quality samples (see one example in the following section) to explore the stellar rotational properties.

\section{Rotational properties of the AMBRE/FEROS stars}
\label{Sec:Discussion}

In this section, we  illustrate  some of the rotational properties of
the AMBRE/FEROS catalogue, by first defining a sub-sample of stars with high-quality derived \Vb. However, we again recall that our derived \Vb\ include the
rotational velocity plus all the other line-broadening processes. In particular, it is affected by the macro-turbulence velocity that is not estimated in the present work. We refer to the discussion of \cite{Fremat2023} about the difficulties to derive \Vma\ from the RVS spectra. 
We also recall that typical \Vma\ values are lower than $\sim$5~\vunits\ in FGKM-type stars, increase with \T, and they are lower for giant stars \citep{Doyle14}. The line-broadening properties that can be derived from our catalogue could 
actually be dominated by macro-turbulence effects for the lowest \Vb\ values (or upper limit ones).

Secondly, it is necessary to investigate if some binaries (or multiple stars) are present in our sample. The \Vb\ derivation for such more complex spectra could indeed not be optimal with our adopted procedure. Moreover, the \Gaia\ identification of the FEROS spectra could be affected by the rather low accuracy of the spectrograph coordinates and could lead to a mismatch in case of multiple stars. For that purpose, we  considered, among the \Gaia\ keywords, {\it duplicated\_source} and {\it non\_single\_star}, which help to identify possible multiple stars.
We found 281 stars in our catalogue with a true value for {\it duplicated\_source}
and 267 with a {\it non\_single\_star} keyword larger than zero.
Finally, our sample of probable single stars ({\it duplicated\_source}=false \& {\it non\_single\_star}=0)  contained 2,067 objects, including three HD stars (among four) that are known to be single objects following CDS/Simbad\footnote{HD 128620 ($\alpha$ Cen A) being a spectroscopist binary has been excluded.}.

Then, we note that the best-quality \Vb\ estimates could be selected by considering only the \Vb\ uncertainties (however, they are not available for upper limit values) since they could be a good indicator of a wrong mask choice (and, hence, low-quality stellar parameters) or low-quality spectra.
However, we favour a more global approach that can be applied to
any star, even those with only upper limits estimates. For that purpose, we considered the $\chi_{\rm AMBRE}^{2}$ quality parameter and the FEROS spectra \SNR\ values. By adopting  $\chi_{\rm AMBRE}^{2}$<-1.5 and \SNR>50 simultaneously to the above-defined single star sample, we selected 1,934 well parameterised stars with an estimated \Vb\ (called the 'High Quality' subsample, hereafter, with this information provided in Table~\ref{Tab:Vbroad}, in the last column). Among them,
we have 1,103 upper limit values and 831 with fully measured \Vb\ values, associated to a mean \Vb\ relative uncertainty smaller than 8\% (there are 806 stars with a \Vb\ relative uncertainty better than 30\%).\\

\begin{figure*}[t]
    \centering
        \includegraphics[width=0.48\linewidth]{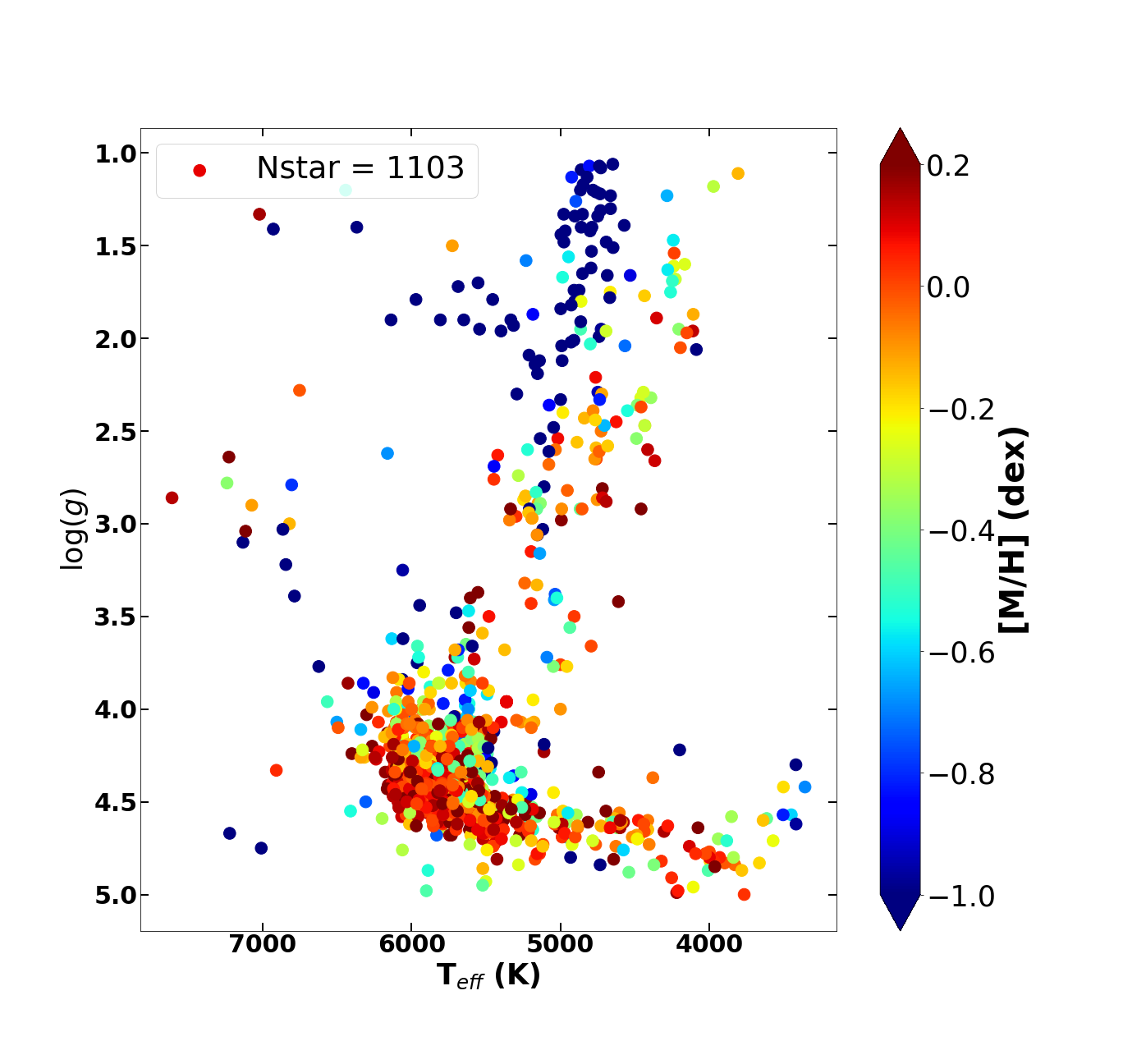}  
    \includegraphics[width=0.48\linewidth]{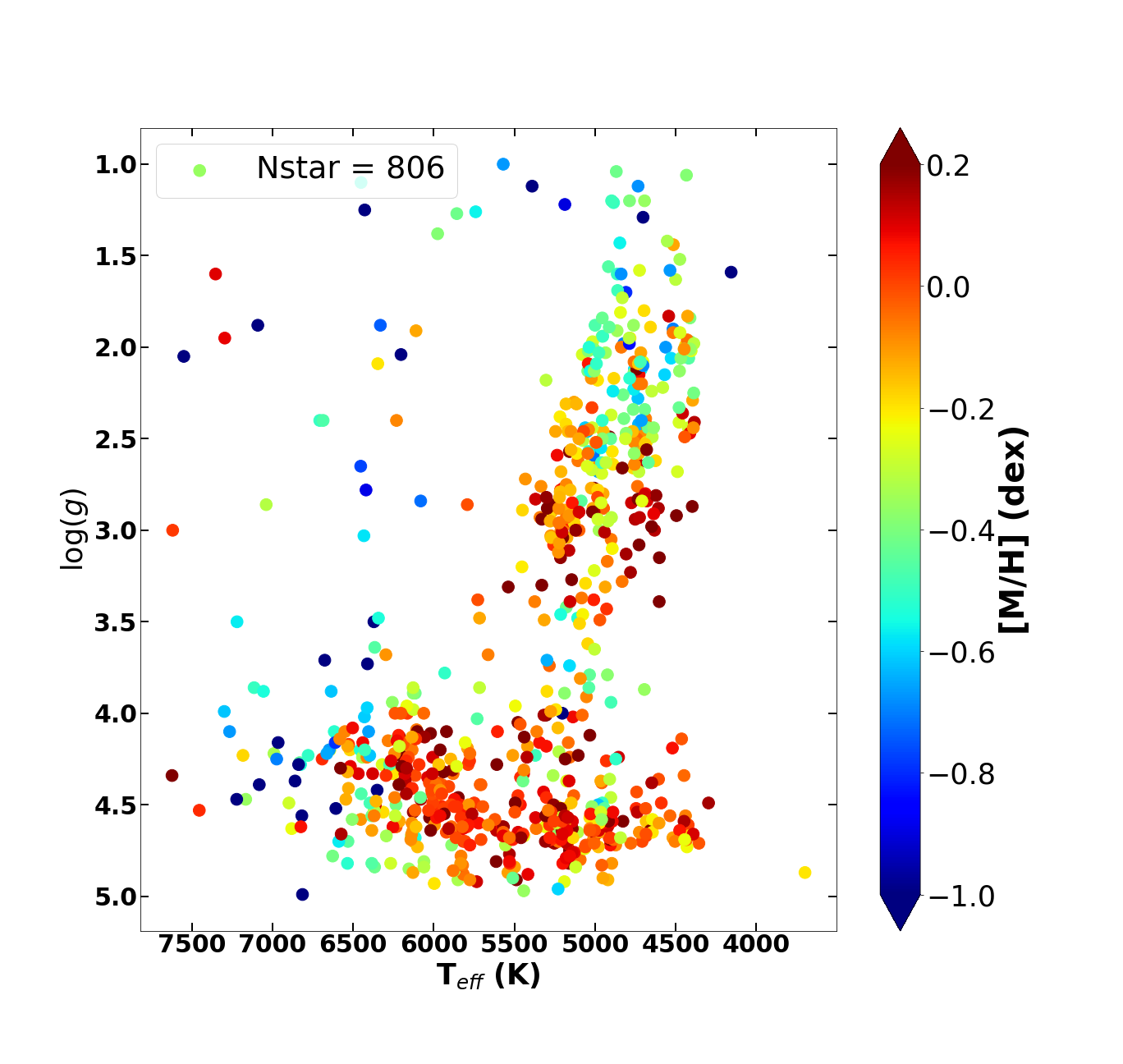}  
    \includegraphics[width=0.48\linewidth]{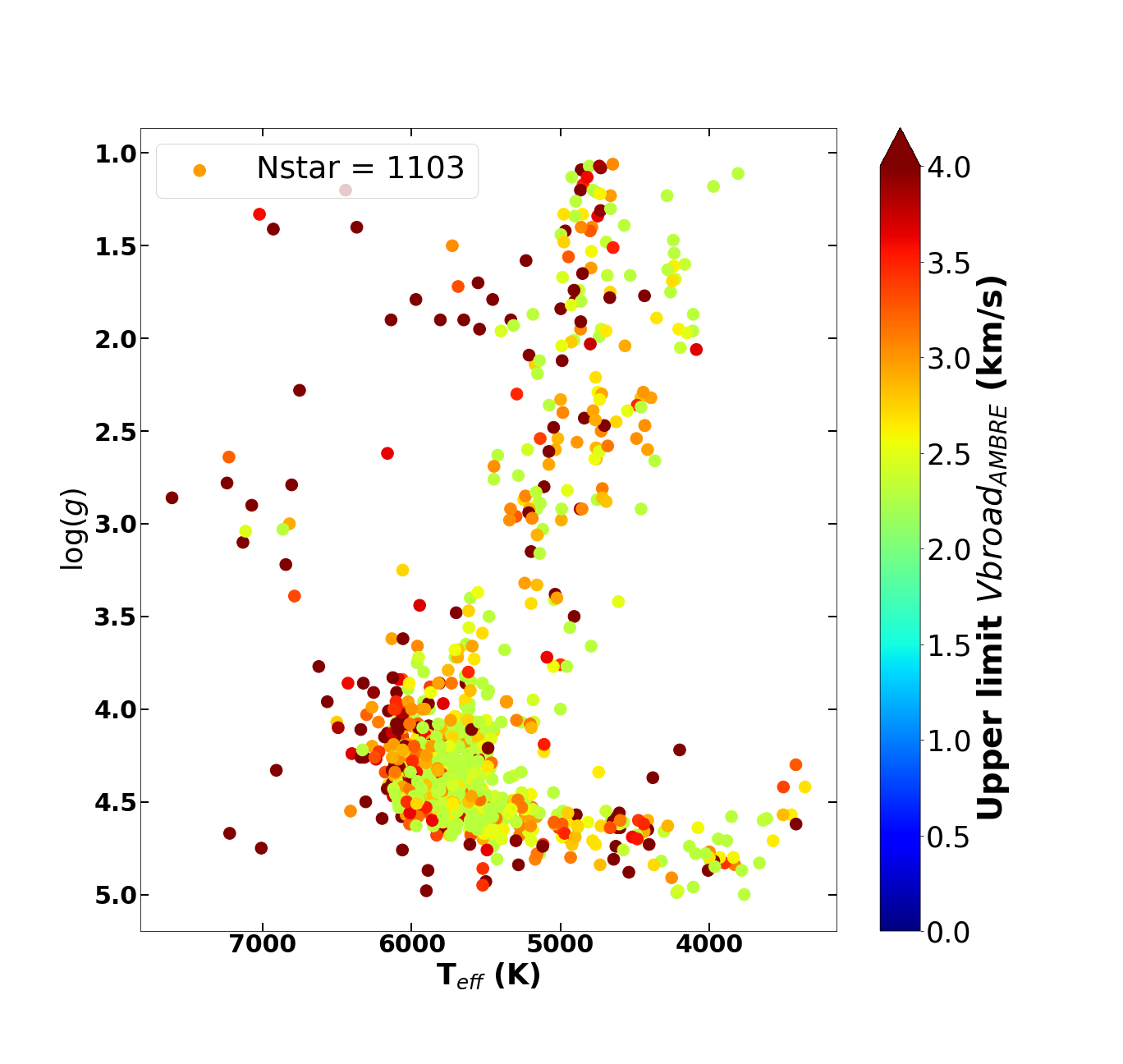}  
    \includegraphics[width=0.48\linewidth]{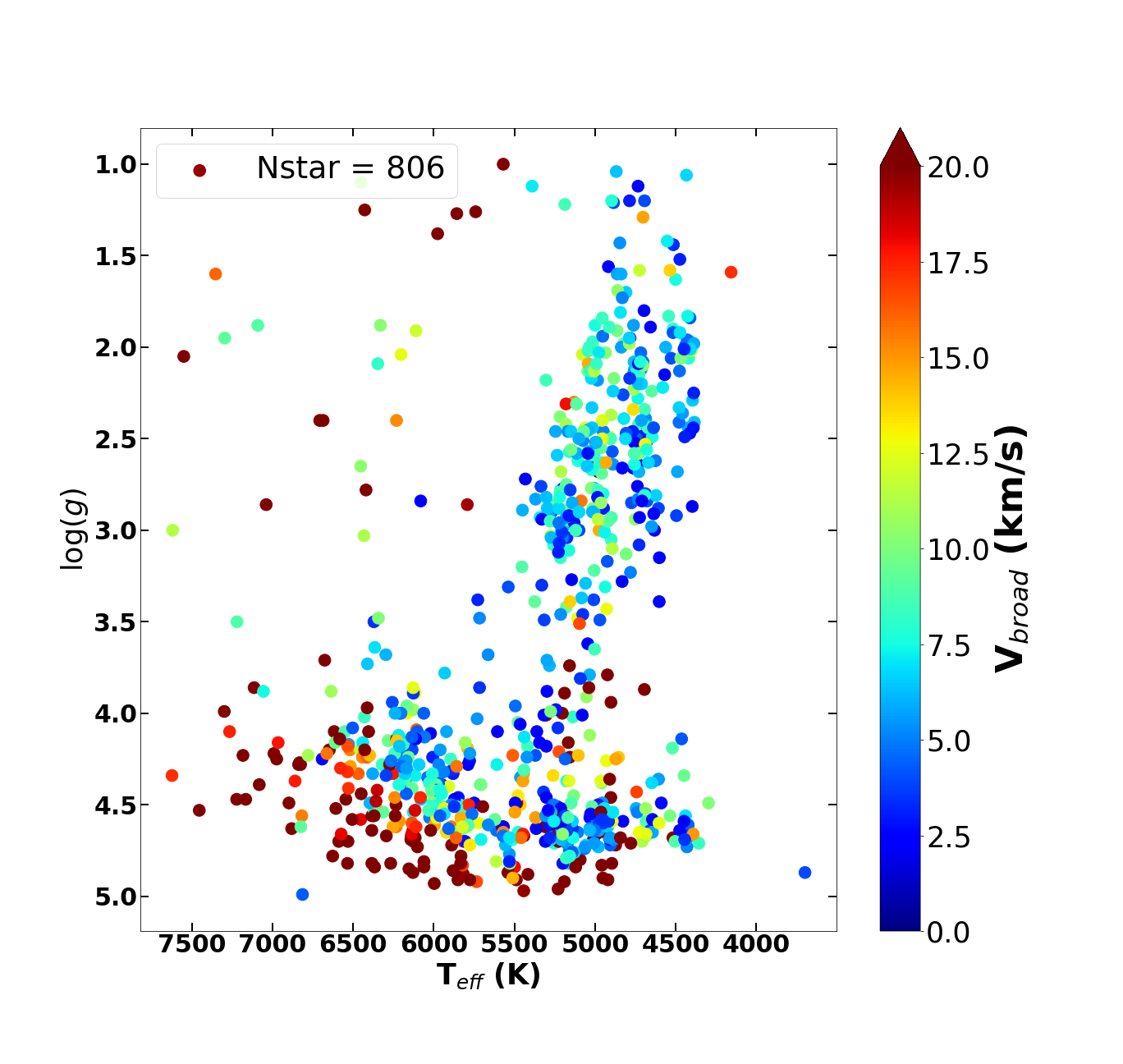}  
    \caption{Kiel diagrams for the AMBRE/FEROS High-Quality subsample. The left and right columns show stars with an upper limit \Vb\ measurement and \Vb\ determined with a relative accuracy better than 30\%, respectively. The stellar metallicity and \Vb\ are adopted for the colour coding (top and bottom, respectively)}
    \label{Fig:Kiel}
\end{figure*}

\begin{figure*}[t]
    \centering
    \includegraphics[width=0.48\linewidth]{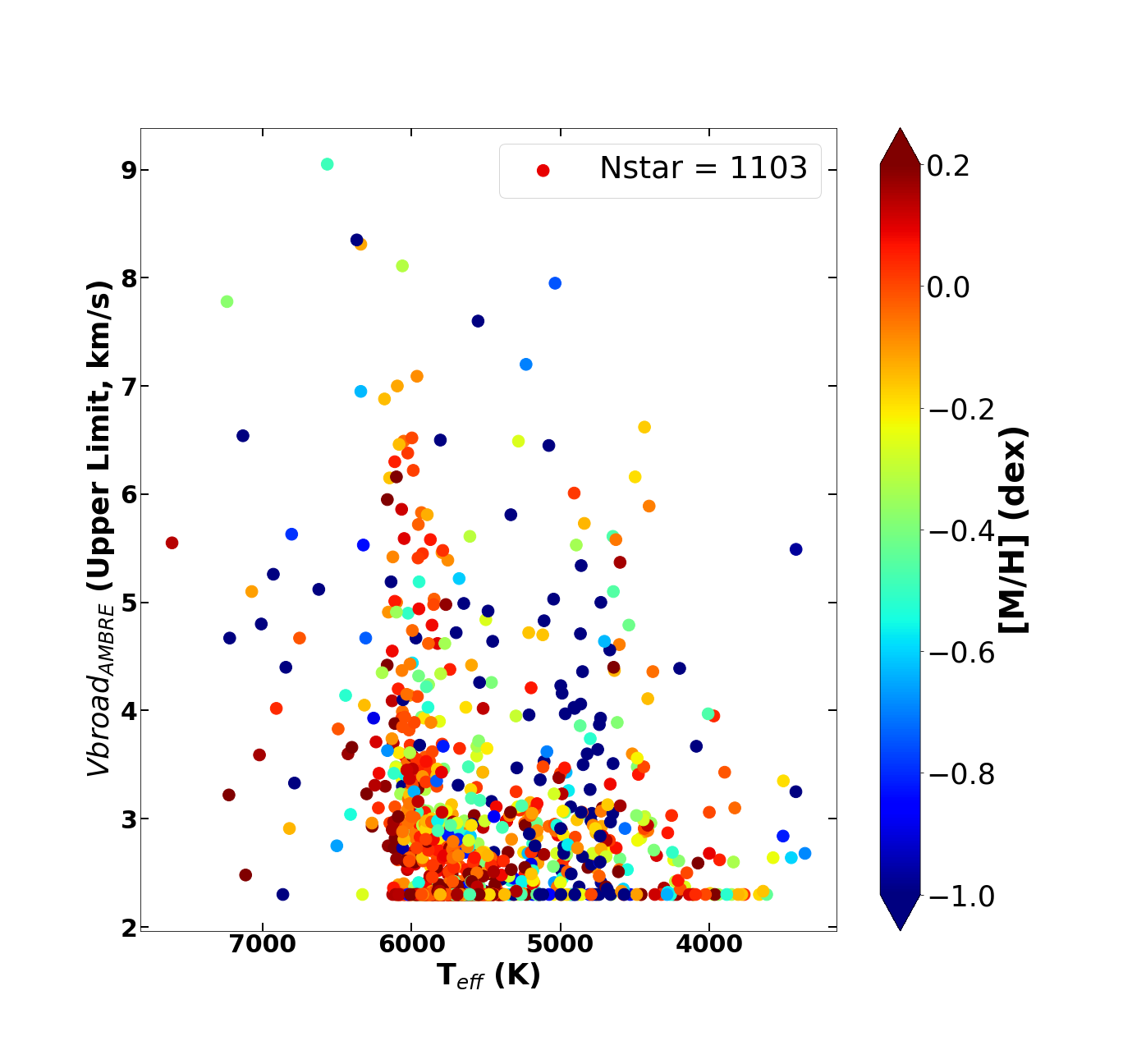}  
    \includegraphics[width=0.48\linewidth]{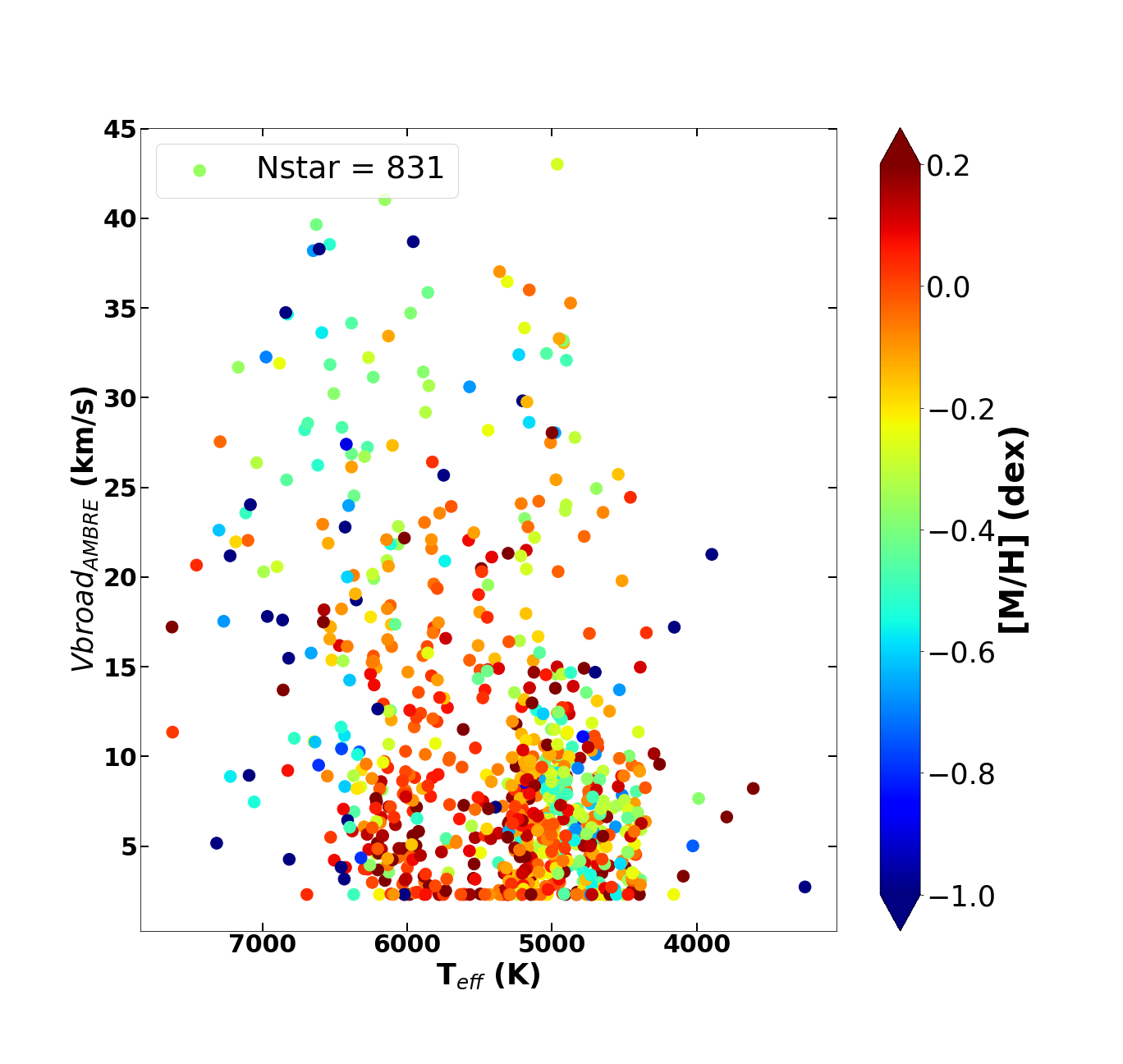}  
    \includegraphics[width=0.48\linewidth]{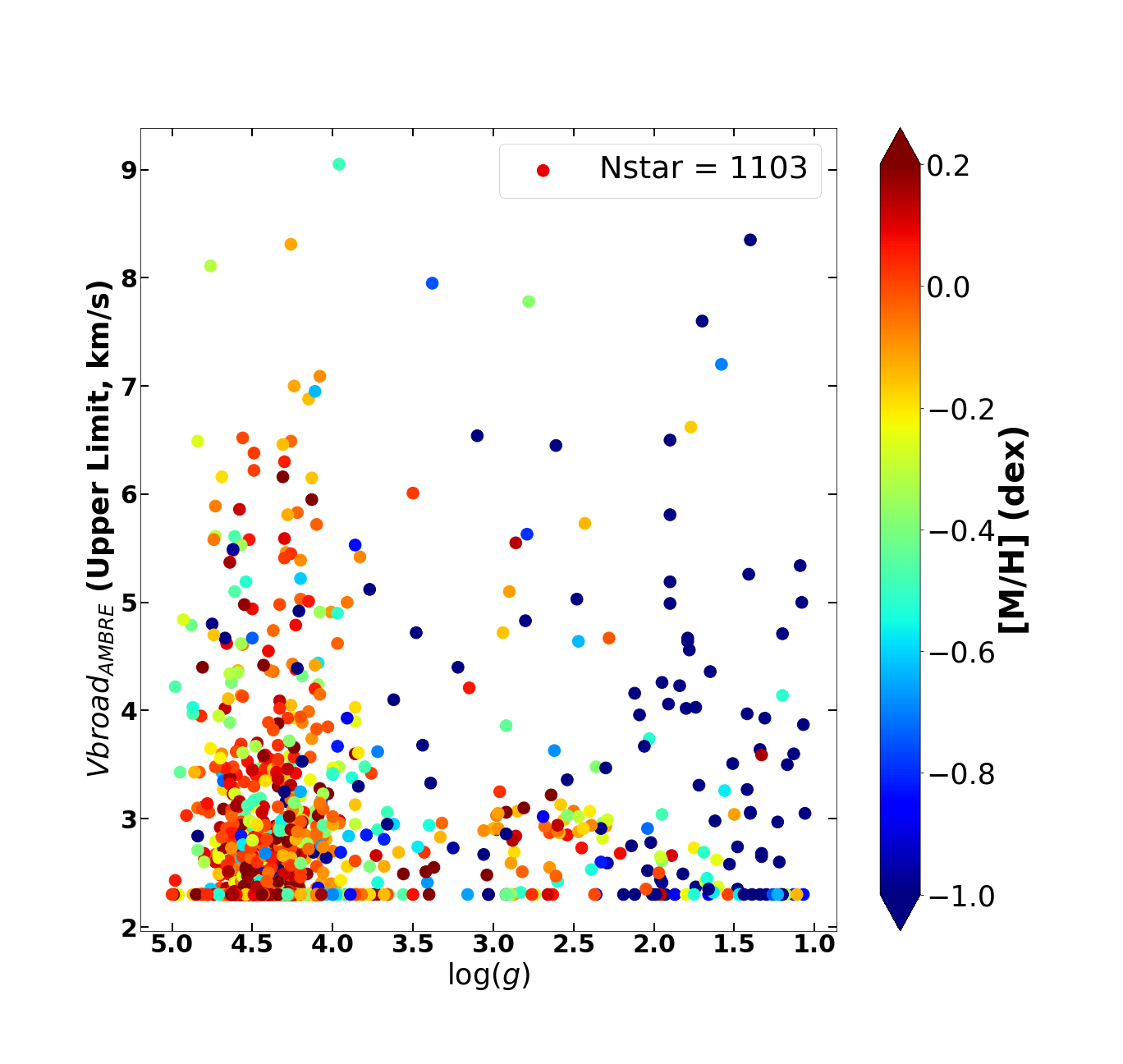}  
    \includegraphics[width=0.48\linewidth]{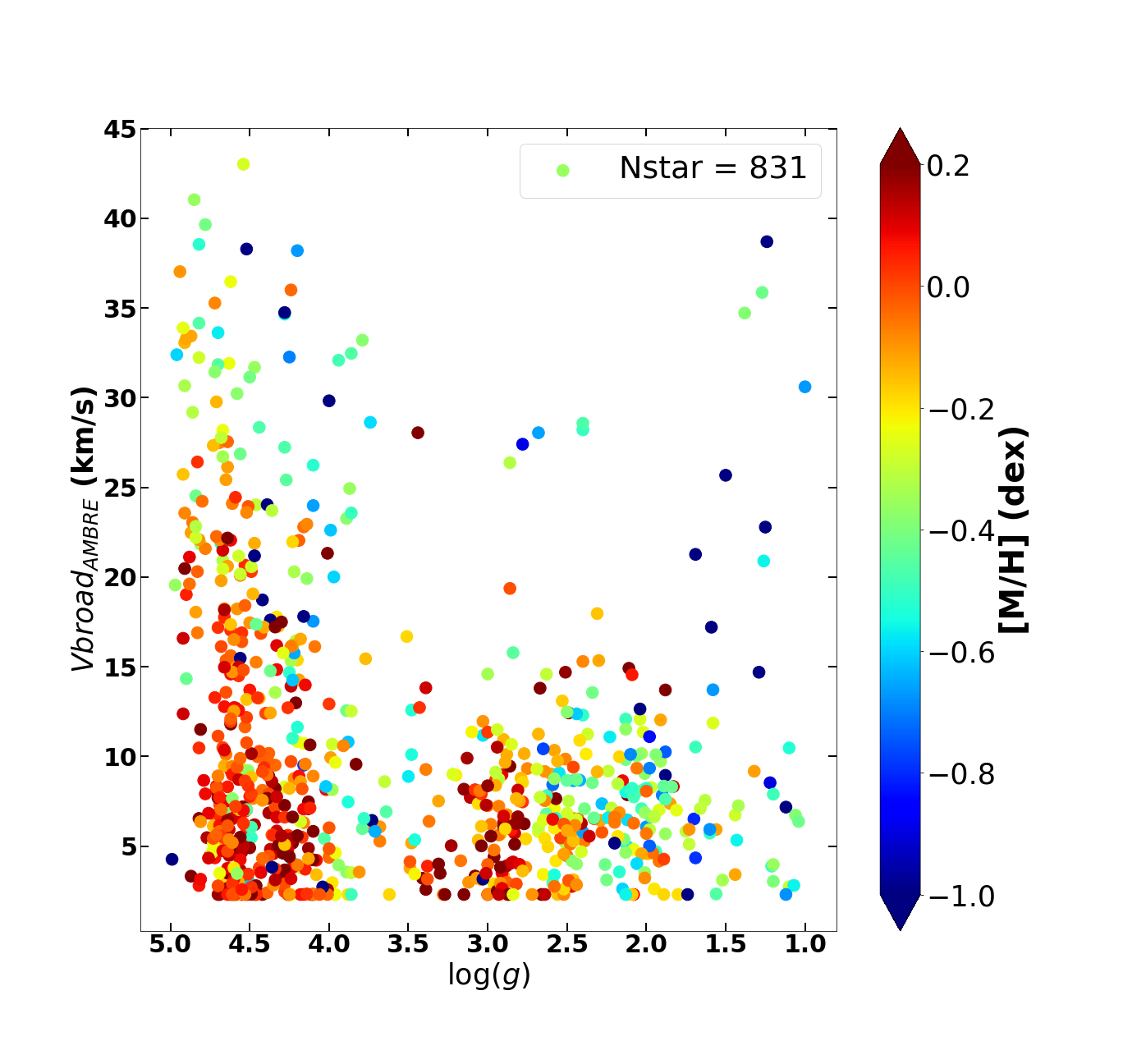}  
    \caption{\Vb\ as a function of the effective temperature (top panels) and the surface gravity (bottom) for the AMBRE/FEROS High-Quality sub-sample. The left and right panels show stars with an upper-limit \Vb\ measurement and \Vb\ determined with a relative accuracy better than 30\%, respectively.}
    \label{Fig:Vb_TeffLogg}
\end{figure*}

To illustrate the properties of this High-Quality subsample,
we show in Fig.~\ref{Fig:Kiel} the Kiel diagrams of the selected
stars colour-coded with their metallicity (top) and broadening velocities (bottom), separating the upper-limits and fully derived \Vb\ values (left and right columns, respectively). It can be seen in these plots that most stellar evolution stages of the FGKM-stellar types have estimated \Vb\ in our catalogue, which also covers a rather broad range of \meta\ values. First, we note that most upper-limit values are smaller than $\sim$5~\vunits,\ which is a typical value of the macro-turbulence velocity for these stellar types (see above). Moreover, several stars with a fully derived \Vb\ (bottom-right panel) have rather low \Vb\ values. It can thus be estimated that most of these stars hardly rotate at all or at extremely low rotational velocities.
Furthermore, the fastest rotators are found for the hottest dwarf stars, as already well known, and again shown in Fig.~\ref{Fig:Vb_TeffLogg}.
These figures illustrate the \Vb\ properties as a function of \T\ and \g, top and bottom panels, respectively. It is seen that most cool dwarfs and giants of our catalogue are predominately slow rotators. In particular, the cool giants characterised by rather low \Vb~values could result from rotational spin-down during post main sequence evolution.
On the contrary, there are few metal-poor supergiants that exhibit larger \Vb~values (bottom-right panel of Fig~\ref{Fig:Vb_TeffLogg}) that could be interpreted as large macro-turbulence
effects and/or preserved main sequence rotation. Among them, we identified five HQ extreme hot supergiants
(\T>5,500~K and \g<1.5) with \Vb>20\vunits. These \Vb\ were measured with an error lower than 2~\vunits\ and their metallicity was found to be  between -1.3 and -0.4~dex (with classical \AF\ enrichment for Galactic stars). 
Three of them are known Cepheid variables with literature \Vsini\ slightly smaller than our \Vb\ (which, however, includes \Vma\ contrarily to \Vsini) and two are suspected post-AGB without any known spectral
information in the literature.
Similarly, we found five HQ dwarf single stars (\T>6,000~K and \g>4.0) in our catalogue that exhibit rather high rotational rates (\Vb>35\vunits, with very low uncertainties)\footnote{Almost 50 similar dwarfs have \Vb>20\vunits.}. Again, they are moderately metal-poor and none of them were previously known as a fast rotator, according to CDS/Simbad.

\section{Summary}
\label{Sec:Conclu}

We  present a catalogue of 2,584 stars whose line-broadening coefficients (a parameter proportional to the stellar rotational velocity) were determined from high spectral resolution FEROS spectra. 
These stars were previously parameterised within the AMBRE Project. They have effective temperature within $\sim$3,500 to $\sim$7,500~K,
stellar surface gravity between 1.0 and 5.0 and any metallicities
between the metal-poor up to sub-Solar regimes.
We recall here that our procedure to estimate the \Vb\ values 
adopts the AMBRE atmospheric parameters, which are assumed to aptly characterise the stellar properties. 
This catalogue is also characterised by its high homogeneity
for the observational data and whole analysis procedure (from the atmospheric
parameter determination up to the line-broadening estimate).

This catalogue has been validated thanks to literature, ground-based spectroscopic surveys, and \Gaia/RVS data.  The agreement with the
comparison samples is found to be good, confirming that the AMBRE/FEROS catalogue of rotational broadening is of high quality. 
 For its use, we recommend considering the relative uncertainties and the different associated quality flags to define high-quality sub-samples to better explore the stellar rotational properties. As an example, by analysing the derived \Vb\ uncertainties, the FEROS \SNR, the AMBRE parameterisation qualities, and the possible binarity contamination thanks to \Gaia\ data, we defined a sub-sample of about 1,900\ stars with high-quality \Vb\ determinations. The associated mean \Vb\ relative uncertainty is smaller than 8\%. 
Checking literature parameterisation and/or spectral determination could also help to select the best \Vb\ determinations. 
It is found that most AMBRE/FEROS stars are rather slow rotators 
although rather extreme \Vb\ values are reported for a few dwarf and supergiant stars,
most of them being unknown as fast rotators and/or affected by large macro-turbulent effects prior to the present work.

Finally, future studies will be devoted to the determination of the line-broadening coefficients of the HARPS and UVES spectra, already parameterised by the AMBRE Project \citep[respectively]{DePascale2014, worley2016}.
This should significantly increase the statistics of rotational information derived from 
high-spectral-resolution spectra for the FGKM stellar types.

\begin{acknowledgements}
      The AMBRE Project was financially supported by ESO, OCA, CNES, ANR, and INSU. PdL and ARB also acknowledge partial funding from the European
Union’s Horizon 2020 research and innovation program under SPACE-H2020
grant agreement number 101004214 (EXPLORE project).

FB acknowledges financial supports for his visits in Nice from the UniCa Erasmus+ International Mobility Program, the French Academy of Sciences, the J.-L. Lagrange Laboratory, the CNES support to the \Gaia\ mission, and the EU H2020 Program.\\

      This work has made use of data from the European Space Agency (ESA)
mission \Gaia\ (https://www.cosmos.esa.int/gaia), processed by the \Gaia\ Data Processing and Analysis Consortium (DPAC, https://www.cosmos.esa.int/web/gaia/dpac/consortium). Funding for the DPAC has been provided by national institutions, in particular the institutions participating in the Gaia Multilateral Agreement.\\

We also used the IPython package \citep{ipython}, NumPy \citep{NumPy}, Matplotlib \citep{Matplotlib}, Pandas, TOPCAT \citep{Topcat} and the SIMBAD database,
operated at CDS, Strasbourg, France \citep{Simbad}. \\

Finally, the authors would like to warmly thank Mamadou N'Diaye and Eric Lagadec who initiated  and made possible this collaboration 
between Ouagadougou and Nice.
\end{acknowledgements}

\bibliographystyle{aa}
\bibliography{ref}

\begin{thebibliography}{52}
\expandafter\ifx\csname natexlab\endcsname\relax\def\natexlab#1{#1}\fi

\bibitem[{{Abdurro'uf} {et~al.}(2022){Abdurro'uf}, {Accetta}, {Aerts}, {Silva Aguirre}, {Ahumada}, {Ajgaonkar}, {Filiz Ak}, {Alam}, {Allende Prieto}, {Almeida}, {Anders}, {Anderson}, {Andrews}, {Anguiano}, {Aquino-Ort{\'\i}z}, {Arag{\'o}n-Salamanca}, {Argudo-Fern{\'a}ndez}, {Ata}, {Aubert}, {Avila-Reese}, {Badenes}, {Barb{\'a}}, {Barger}, {Barrera-Ballesteros}, {Beaton}, {Beers}, {Belfiore}, {Bender}, {Bernardi}, {Bershady}, {Beutler}, {Bidin}, {Bird}, {Bizyaev}, {Blanc}, {Blanton}, {Boardman}, {Bolton}, {Boquien}, {Borissova}, {Bovy}, {Brandt}, {Brown}, {Brownstein}, {Brusa}, {Buchner}, {Bundy}, {Burchett}, {Bureau}, {Burgasser}, {Cabang}, {Campbell}, {Cappellari}, {Carlberg}, {Wanderley}, {Carrera}, {Cash}, {Chen}, {Chen}, {Cherinka}, {Chiappini}, {Choi}, {Chojnowski}, {Chung}, {Clerc}, {Cohen}, {Comerford}, {Comparat}, {da Costa}, {Covey}, {Crane}, {Cruz-Gonzalez}, {Culhane}, {Cunha}, {Dai}, {Damke}, {Darling}, {Davidson}, {Davies}, {Dawson}, {De Lee}, {Diamond-Stanic}, {Cano-D{\'\i}az}, {S{\'a}nchez},
  {Donor}, {Duckworth}, {Dwelly}, {Eisenstein}, {Elsworth}, {Emsellem}, {Eracleous}, {Escoffier}, {Fan}, {Farr}, {Feng}, {Fern{\'a}ndez-Trincado}, {Feuillet}, {Filipp}, {Fillingham}, {Frinchaboy}, {Fromenteau}, {Galbany}, {Garc{\'\i}a}, {Garc{\'\i}a-Hern{\'a}ndez}, {Ge}, {Geisler}, {Gelfand}, {G{\'e}ron}, {Gibson}, {Goddy}, {Godoy-Rivera}, {Grabowski}, {Green}, {Greener}, {Grier}, {Griffith}, {Guo}, {Guy}, {Hadjara}, {Harding}, {Hasselquist}, {Hayes}, {Hearty}, {Hern{\'a}ndez}, {Hill}, {Hogg}, {Holtzman}, {Horta}, {Hsieh}, {Hsu}, {Hsu}, {Huber}, {Huertas-Company}, {Hutchinson}, {Hwang}, {Ibarra-Medel}, {Chitham}, {Ilha}, {Imig}, {Jaekle}, {Jayasinghe}, {Ji}, {Johnson}, {Jones}, {J{\"o}nsson}, {Katkov}, {Khalatyan}, {Kinemuchi}, {Kisku}, {Knapen}, {Kneib}, {Kollmeier}, {Kong}, {Kounkel}, {Kreckel}, {Krishnarao}, {Lacerna}, {Lane}, {Langgin}, {Lavender}, {Law}, {Lazarz}, {Leung}, {Leung}, {Lewis}, {Li}, {Li}, {Lian}, {Liang}, {Lin}, {Lin}, {Lin}, {Lintott}, {Long}, {Longa-Pe{\~n}a}, {L{\'o}pez-Cob{\'a}}, {Lu},
  {Lundgren}, {Luo}, {Mackereth}, {de la Macorra}, {Mahadevan}, {Majewski}, {Manchado}, {Mandeville}, {Maraston}, {Margalef-Bentabol}, {Masseron}, {Masters}, {Mathur}, {McDermid}, {Mckay}, {Merloni}, {Merrifield}, {Meszaros}, {Miglio}, {Di Mille}, {Minniti}, {Minsley}, \& {Monachesi}}]{Abdurrouf2022}
{Abdurro'uf}, {Accetta}, K., {Aerts}, C., {et~al.} 2022, \apjs, 259, 35

\bibitem[{{Adibekyan} {et~al.}(2015){Adibekyan}, {Figueira}, {Santos}, {Sousa}, {Faria}, {Delgado-Mena}, {Oshagh}, {Tsantaki}, {Hakobyan}, {Gonz{\'a}lez Hern{\'a}ndez}, {Su{\'a}rez-Andr{\'e}s}, \& {Israelian}}]{Adibekyan2015}
{Adibekyan}, V., {Figueira}, P., {Santos}, N.~C., {et~al.} 2015, \aap, 583, A94

\bibitem[{Angus {et~al.}(2015)Angus, Aigrain, Foreman-Mackey, \& McQuillan}]{Angus2015}
Angus, R., Aigrain, S., Foreman-Mackey, D., \& McQuillan, A. 2015, Monthly Notices of the Royal Astronomical Society, 450, 1787

\bibitem[{{Baranne} {et~al.}(1979){Baranne}, {Mayor}, \& {Poncet}}]{Baranne1979}
{Baranne}, A., {Mayor}, M., \& {Poncet}, J.~L. 1979, Vistas in Astronomy, 23, 279

\bibitem[{Barnes(2007)}]{Barnes2007}
Barnes, S.~A. 2007, The Astrophysical Journal, 669, 1167

\bibitem[{{Bragan{\c{c}}a} {et~al.}(2012){Bragan{\c{c}}a}, {Daflon}, {Cunha}, {Bensby}, {Oey}, \& {Walth}}]{Bragan12}
{Bragan{\c{c}}a}, G.~A., {Daflon}, S., {Cunha}, K., {et~al.} 2012, \aj, 144, 130

\bibitem[{{Buder} {et~al.}(2025){Buder}, {Kos}, {Wang}, {McKenzie}, {Howell}, {Martell}, {Hayden}, {Zucker}, {Nordlander}, {Montet}, {Traven}, {Bland-Hawthorn}, {de Silva}, {Freeman}, {Lewis}, {Lind}, {Sharma}, {Simpson}, {Stello}, {Zwitter}, {Amarsi}, {Armstrong}, {Banks}, {Beavis}, {Beeson}, {Chen}, {Ciuc{\u{a}}}, {da Costa}, {de Grijs}, {Martin}, {Nataf}, {Ness}, {Rains}, {Scarr}, {Vogrin{\v{c}}i{\v{c}}}, {Wang}, {Wittenmyer}, {Xie}, \& {The Galah Collaboration}}]{Buder2024}
{Buder}, S., {Kos}, J., {Wang}, X.~E., {et~al.} 2025, \pasa, 42, e051

\bibitem[{Chiappini {et~al.}(2003)Chiappini, Matteucci, \& Meynet}]{Chiappini2003}
Chiappini, C., Matteucci, F., \& Meynet, G. 2003, Astronomy \& Astrophysics, 410, 257

\bibitem[{{Claret}(2000)}]{Claret2000}
{Claret}, A. 2000, \aap, 363, 1081

\bibitem[{Contursi {et~al.}(2024)Contursi, de~Laverny, Recio-Blanco, Palicio, \& Abia}]{Contursi2024}
Contursi, G., de~Laverny, P., Recio-Blanco, A., Palicio, P., \& Abia, C. 2024, Astronomy \& Astrophysics, 683, A138

\bibitem[{{Costa-Almeida} {et~al.}(2021){Costa-Almeida}, {Porto de Mello}, {Giribaldi}, {Lorenzo-Oliveira}, \& {Ubaldo-Melo}}]{Costa21}
{Costa-Almeida}, E., {Porto de Mello}, G.~F., {Giribaldi}, R.~E., {Lorenzo-Oliveira}, D., \& {Ubaldo-Melo}, M.~L. 2021, \mnras, 508, 5148

\bibitem[{{Cropper} {et~al.}(2018){Cropper}, {Katz}, {Sartoretti}, {Prusti}, {de Bruijne}, {Chassat}, {Charvet}, {Boyadjian}, {Perryman}, {Sarri}, {Gare}, {Erdmann}, {Munari}, {Zwitter}, {Wilkinson}, {Arenou}, {Vallenari}, {G{\'o}mez}, {Panuzzo}, {Seabroke}, {Allende Prieto}, {Benson}, {Marchal}, {Huckle}, {Smith}, {Dolding}, {Jan{\ss}en}, {Viala}, {Blomme}, {Baker}, {Boudreault}, {Crifo}, {Soubiran}, {Fr{\'e}mat}, {Jasniewicz}, {Guerrier}, {Guy}, {Turon}, {Jean-Antoine-Piccolo}, {Th{\'e}venin}, {David}, {Gosset}, \& {Damerdji}}]{RVS18}
{Cropper}, M., {Katz}, D., {Sartoretti}, P., {et~al.} 2018, \aap, 616, A5

\bibitem[{{de Burgos} {et~al.}(2025){de Burgos}, {Sim{\'o}n-D{\'\i}az}, {Urbaneja}, {Holgado}, {Ekstr{\"o}m}, {Ram{\'\i}rez-Tannus}, \& {Zari}}]{Burgos25}
{de Burgos}, A., {Sim{\'o}n-D{\'\i}az}, S., {Urbaneja}, M.~A., {et~al.} 2025, \aap, 695, A87

\bibitem[{{de Laverny} {et~al.}(2013){de Laverny}, {Recio-Blanco}, {Worley}, {De Pascale}, {Hill}, \& {Bijaoui}}]{deLaverny2013}
{de Laverny}, P., {Recio-Blanco}, A., {Worley}, C.~C., {et~al.} 2013, The Messenger, 153, 18

\bibitem[{{de Laverny} {et~al.}(2012){de Laverny}, {Recio-Blanco}, {Worley}, \& {Plez}}]{deLaverny2012}
{de Laverny}, P., {Recio-Blanco}, A., {Worley}, C.~C., \& {Plez}, B. 2012, \aap, 544, A126

\bibitem[{De~Medeiros {et~al.}(2006)De~Medeiros, Silva, Do~Nascimento~Jr, Martins, da~Silva, Melo, \& Burnet}]{JRM2006}
De~Medeiros, J., Silva, J., Do~Nascimento~Jr, J., {et~al.} 2006, Astronomy \& Astrophysics, 458, 895

\bibitem[{{de Medeiros} \& {Mayor}(1999)}]{JRM1999}
{de Medeiros}, J.~R. \& {Mayor}, M. 1999, \aaps, 139, 433

\bibitem[{{De Medeiros} {et~al.}(2002){De Medeiros}, {Udry}, {Burki}, \& {Mayor}}]{JRM2002}
{De Medeiros}, J.~R., {Udry}, S., {Burki}, G., \& {Mayor}, M. 2002, \aap, 395, 97

\bibitem[{{De Medeiros} {et~al.}(2004){De Medeiros}, {Udry}, \& {Mayor}}]{JRM2004}
{De Medeiros}, J.~R., {Udry}, S., \& {Mayor}, M. 2004, \aap, 427, 313

\bibitem[{{De Pascale} {et~al.}(2014){De Pascale}, {Worley}, {de Laverny}, {Recio-Blanco}, {Hill}, \& {Bijaoui}}]{DePascale2014}
{De Pascale}, M., {Worley}, C.~C., {de Laverny}, P., {et~al.} 2014, \aap, 570, A68

\bibitem[{{De Silva} {et~al.}(2015){De Silva}, {Freeman}, {Bland-Hawthorn}, {Martell}, {de Boer}, {Asplund}, {Keller}, {Sharma}, {Zucker}, {Zwitter}, {Anguiano}, {Bacigalupo}, {Bayliss}, {Beavis}, {Bergemann}, {Campbell}, {Cannon}, {Carollo}, {Casagrande}, {Casey}, {Da Costa}, {D'Orazi}, {Dotter}, {Duong}, {Heger}, {Ireland}, {Kafle}, {Kos}, {Lattanzio}, {Lewis}, {Lin}, {Lind}, {Munari}, {Nataf}, {O'Toole}, {Parker}, {Reid}, {Schlesinger}, {Sheinis}, {Simpson}, {Stello}, {Ting}, {Traven}, {Watson}, {Wittenmyer}, {Yong}, \& {{\v{Z}}erjal}}]{De_Silva2015}
{De Silva}, G.~M., {Freeman}, K.~C., {Bland-Hawthorn}, J., {et~al.} 2015, \mnras, 449, 2604

\bibitem[{{Doyle} {et~al.}(2014){Doyle}, {Davies}, {Smalley}, {Chaplin}, \& {Elsworth}}]{Doyle14}
{Doyle}, A.~P., {Davies}, G.~R., {Smalley}, B., {Chaplin}, W.~J., \& {Elsworth}, Y. 2014, \mnras, 444, 3592

\bibitem[{{Fr{\'e}mat} {et~al.}(2023){Fr{\'e}mat}, {Royer}, {Marchal}, {Blomme}, {Sartoretti}, {Guerrier}, {Panuzzo}, {Katz}, {Seabroke}, {Th{\'e}venin}, {Cropper}, {Benson}, {Damerdji}, {Haigron}, {Lobel}, {Smith}, {Baker}, {Chemin}, {David}, {Dolding}, {Gosset}, {Jan{\ss}en}, {Jasniewicz}, {Plum}, {Samaras}, {Snaith}, {Soubiran}, {Vanel}, {Zorec}, {Zwitter}, {Brouillet}, {Caffau}, {Crifo}, {Fabre}, {Fragkoudi}, {Huckle}, {Lasne}, {Leclerc}, {Mastrobuono-Battisti}, {Jean-Antoine Piccolo}, \& {Viala}}]{Fremat2023}
{Fr{\'e}mat}, Y., {Royer}, F., {Marchal}, O., {et~al.} 2023, \aap, 674, A8

\bibitem[{{Gaia Collaboration} {et~al.}(2016){Gaia Collaboration}, {Prusti}, {de Bruijne}, {Brown}, {Vallenari}, {Babusiaux}, {Bailer-Jones}, {Bastian}, {Biermann}, {Evans}, {Eyer}, {Jansen}, {Jordi}, {Klioner}, {Lammers}, {Lindegren}, {Luri}, {Mignard}, {Milligan}, {Panem}, {Poinsignon}, {Pourbaix}, {Randich}, {Sarri}, {Sartoretti}, {Siddiqui}, {Soubiran}, {Valette}, {van Leeuwen}, {Walton}, {Aerts}, {Arenou}, {Cropper}, {Drimmel}, {H{\o}g}, {Katz}, {Lattanzi}, {O'Mullane}, {Grebel}, {Holland}, {Huc}, {Passot}, {Bramante}, {Cacciari}, {Casta{\~n}eda}, {Chaoul}, {Cheek}, {De Angeli}, {Fabricius}, {Guerra}, {Hern{\'a}ndez}, {Jean-Antoine-Piccolo}, {Masana}, {Messineo}, {Mowlavi}, {Nienartowicz}, {Ord{\'o}{\~n}ez-Blanco}, {Panuzzo}, {Portell}, {Richards}, {Riello}, {Seabroke}, {Tanga}, {Th{\'e}venin}, {Torra}, {Els}, {Gracia-Abril}, {Comoretto}, {Garcia-Reinaldos}, {Lock}, {Mercier}, {Altmann}, {Andrae}, {Astraatmadja}, {Bellas-Velidis}, {Benson}, {Berthier}, {Blomme}, {Busso}, {Carry}, {Cellino}, {Clementini},
  {Cowell}, {Creevey}, {Cuypers}, {Davidson}, {De Ridder}, {de Torres}, {Delchambre}, {Dell'Oro}, {Ducourant}, {Fr{\'e}mat}, {Garc{\'\i}a-Torres}, {Gosset}, {Halbwachs}, {Hambly}, {Harrison}, {Hauser}, {Hestroffer}, {Hodgkin}, {Huckle}, {Hutton}, {Jasniewicz}, {Jordan}, {Kontizas}, {Korn}, {Lanzafame}, {Manteiga}, {Moitinho}, {Muinonen}, {Osinde}, {Pancino}, {Pauwels}, {Petit}, {Recio-Blanco}, {Robin}, {Sarro}, {Siopis}, {Smith}, {Smith}, {Sozzetti}, {Thuillot}, {van Reeven}, {Viala}, {Abbas}, {Abreu Aramburu}, {Accart}, {Aguado}, {Allan}, {Allasia}, {Altavilla}, {{\'A}lvarez}, {Alves}, {Anderson}, {Andrei}, {Anglada Varela}, {Antiche}, {Antoja}, {Ant{\'o}n}, {Arcay}, {Atzei}, {Ayache}, {Bach}, {Baker}, {Balaguer-N{\'u}{\~n}ez}, {Barache}, {Barata}, {Barbier}, {Barblan}, {Baroni}, {Barrado y Navascu{\'e}s}, {Barros}, {Barstow}, {Becciani}, {Bellazzini}, {Bellei}, {Bello Garc{\'\i}a}, {Belokurov}, {Bendjoya}, {Berihuete}, {Bianchi}, {Bienaym{\'e}}, {Billebaud}, {Blagorodnova}, {Blanco-Cuaresma}, {Boch},
  {Bombrun}, {Borrachero}, {Bouquillon}, {Bourda}, {Bouy}, {Bragaglia}, {Breddels}, {Brouillet}, {Br{\"u}semeister}, {Bucciarelli}, {Budnik}, {Burgess}, {Burgon}, {Burlacu}, {Busonero}, {Buzzi}, {Caffau}, {Cambras}, {Campbell}, {Cancelliere}, {Cantat-Gaudin}, {Carlucci}, {Carrasco}, {Castellani}, {Charlot}, {Charnas}, {Charvet}, {Chassat}, {Chiavassa}, {Clotet}, {Cocozza}, {Collins}, {Collins}, \& {Costigan}}]{Gaia_Collaboration2016}
{Gaia Collaboration}, {Prusti}, T., {de Bruijne}, J.~H.~J., {et~al.} 2016, \aap, 595, A1

\bibitem[{{Gilmore} {et~al.}(2012){Gilmore}, {Randich}, {Asplund}, {Binney}, {Bonifacio}, {Drew}, {Feltzing}, {Ferguson}, {Jeffries}, {Micela}, {Negueruela}, {Prusti}, {Rix}, {Vallenari}, {Alfaro}, {Allende-Prieto}, {Babusiaux}, {Bensby}, {Blomme}, {Bragaglia}, {Flaccomio}, {Fran{\c{c}}ois}, {Irwin}, {Koposov}, {Korn}, {Lanzafame}, {Pancino}, {Paunzen}, {Recio-Blanco}, {Sacco}, {Smiljanic}, {Van Eck}, {Walton}, {Aden}, {Aerts}, {Affer}, {Alcala}, {Altavilla}, {Alves}, {Antoja}, {Arenou}, {Argiroffi}, {Asensio Ramos}, {Bailer-Jones}, {Balaguer-Nunez}, {Bayo}, {Barbuy}, {Barisevicius}, {Barrado y Navascues}, \& {Mowlavi}}]{Gilmore2012}
{Gilmore}, G., {Randich}, S., {Asplund}, M., {et~al.} 2012, The Messenger, 147, 25

\bibitem[{Glebocki \& Gnacinski(2005)}]{GG2005}
Glebocki, R. \& Gnacinski, P. 2005, VizieR online data catalog, III

\bibitem[{Gray(2008)}]{Gray2008}
Gray, D. 2008, The observation and analysis of stellar photosphere (Cambridge university press)

\bibitem[{Harris {et~al.}(2020)Harris, Millman, van~der Walt, Gommers, Virtanen, Cournapeau, Wieser, Taylor, Berg, Smith, Kern, Picus, Hoyer, van Kerkwijk, Brett, Haldane, del R{'{\i}}o, Wiebe, Peterson, G{'{e}}rard-Marchant, Sheppard, Reddy, Weckesser, Abbasi, Gohlke, \& Oliphant}]{NumPy}
Harris, C.~R., Millman, K.~J., van~der Walt, S.~J., {et~al.} 2020, Nature, 585, 357

\bibitem[{{Holgado} {et~al.}(2022){Holgado}, {Sim{\'o}n-D{\'\i}az}, {Herrero}, \& {Barb{\'a}}}]{Holgado22}
{Holgado}, G., {Sim{\'o}n-D{\'\i}az}, S., {Herrero}, A., \& {Barb{\'a}}, R.~H. 2022, \aap, 665, A150

\bibitem[{Hunter(2007)}]{Matplotlib}
Hunter, J.~D. 2007, Computing In Science \& Engineering, 9, 90

\bibitem[{{Kaufer} {et~al.}(1999){Kaufer}, {Stahl}, {Tubbesing}, {N{\o}rregaard}, {Avila}, {Francois}, {Pasquini}, \& {Pizzella}}]{Kaufer1999}
{Kaufer}, A., {Stahl}, O., {Tubbesing}, S., {et~al.} 1999, The Messenger, 95, 8

\bibitem[{Maeder \& Meynet(2000)}]{Maeder2000}
Maeder, A. \& Meynet, G. 2000, Annual Review of Astronomy and Astrophysics, 38, 143

\bibitem[{{Majewski} {et~al.}(2017){Majewski}, {Schiavon}, {Frinchaboy}, {Allende Prieto}, {Barkhouser}, {Bizyaev}, {Blank}, {Brunner}, {Burton}, {Carrera}, {Chojnowski}, {Cunha}, {Epstein}, {Fitzgerald}, {Garc{\'\i}a P{\'e}rez}, {Hearty}, {Henderson}, {Holtzman}, {Johnson}, {Lam}, {Lawler}, {Maseman}, {M{\'e}sz{\'a}ros}, {Nelson}, {Nguyen}, {Nidever}, {Pinsonneault}, {Shetrone}, {Smee}, {Smith}, {Stolberg}, {Skrutskie}, {Walker}, {Wilson}, {Zasowski}, {Anders}, {Basu}, {Beland}, {Blanton}, {Bovy}, {Brownstein}, {Carlberg}, {Chaplin}, {Chiappini}, {Eisenstein}, {Elsworth}, {Feuillet}, {Fleming}, {Galbraith-Frew}, {Garc{\'\i}a}, {Garc{\'\i}a-Hern{\'a}ndez}, {Gillespie}, {Girardi}, {Gunn}, {Hasselquist}, {Hayden}, {Hekker}, {Ivans}, {Kinemuchi}, {Klaene}, {Mahadevan}, {Mathur}, {Mosser}, {Muna}, {Munn}, {Nichol}, {O'Connell}, {Parejko}, {Robin}, {Rocha-Pinto}, {Schultheis}, {Serenelli}, {Shane}, {Silva Aguirre}, {Sobeck}, {Thompson}, {Troup}, {Weinberg}, \& {Zamora}}]{Majewski2017}
{Majewski}, S.~R., {Schiavon}, R.~P., {Frinchaboy}, P.~M., {et~al.} 2017, \aj, 154, 94

\bibitem[{Melo {et~al.}(2001)Melo, Pasquini, \& De~Medeiros}]{Melo2001}
Melo, C., Pasquini, L., \& De~Medeiros, J. 2001, Astronomy \& Astrophysics, 375, 851

\bibitem[{Perdigon {et~al.}(2021)Perdigon, De~Laverny, Recio-Blanco, Fernandez-Alvar, Santos-Peral, Kordopatis, \& {\'A}lvarez}]{Perdigon2021}
Perdigon, J., De~Laverny, P., Recio-Blanco, A., {et~al.} 2021, Astronomy \& Astrophysics, 647, A162

\bibitem[{P\'erez \& Granger(2007)}]{ipython}
P\'erez, F. \& Granger, B.~E. 2007, Computing in Science and Engineering, 9, 21

\bibitem[{{Prantzos} {et~al.}(2020){Prantzos}, {Abia}, {Cristallo}, {Limongi}, \& {Chieffi}}]{Prantzos20}
{Prantzos}, N., {Abia}, C., {Cristallo}, S., {Limongi}, M., \& {Chieffi}, A. 2020, \mnras, 491, 1832

\bibitem[{{Prantzos} {et~al.}(2018){Prantzos}, {Abia}, {Limongi}, {Chieffi}, \& {Cristallo}}]{Prantzos18}
{Prantzos}, N., {Abia}, C., {Limongi}, M., {Chieffi}, A., \& {Cristallo}, S. 2018, \mnras, 476, 3432

\bibitem[{{Recio-Blanco} {et~al.}(2023){Recio-Blanco}, {de Laverny}, {Palicio}, {Kordopatis}, {{\'A}lvarez}, {Schultheis}, {Contursi}, {Zhao}, {Torralba Elipe}, {Ordenovic}, {Manteiga}, {Dafonte}, {Oreshina-Slezak}, {Bijaoui}, {Fr{\'e}mat}, {Seabroke}, {Pailler}, {Spitoni}, {Poggio}, {Creevey}, {Abreu Aramburu}, {Accart}, {Andrae}, {Bailer-Jones}, {Bellas-Velidis}, {Brouillet}, {Brugaletta}, {Burlacu}, {Carballo}, {Casamiquela}, {Chiavassa}, {Cooper}, {Dapergolas}, {Delchambre}, {Dharmawardena}, {Drimmel}, {Edvardsson}, {Fouesneau}, {Garabato}, {Garc{\'\i}a-Lario}, {Garc{\'\i}a-Torres}, {Gavel}, {Gomez}, {Gonz{\'a}lez-Santamar{\'\i}a}, {Hatzidimitriou}, {Heiter}, {Jean-Antoine Piccolo}, {Kontizas}, {Korn}, {Lanzafame}, {Lebreton}, {Le Fustec}, {Licata}, {Lindstr{\o}m}, {Livanou}, {Lobel}, {Lorca}, {Magdaleno Romeo}, {Marocco}, {Marshall}, {Mary}, {Nicolas}, {Pallas-Quintela}, {Panem}, {Pichon}, {Riclet}, {Robin}, {Rybizki}, {Santove{\~n}a}, {Silvelo}, {Smart}, {Sarro}, {Sordo}, {Soubiran}, {S{\"u}veges},
  {Ulla}, {Vallenari}, {Zorec}, {Utrilla}, \& {Bakker}}]{Recio2023}
{Recio-Blanco}, A., {de Laverny}, P., {Palicio}, P.~A., {et~al.} 2023, \aap, 674, A29

\bibitem[{{Recio-Blanco} {et~al.}(2002){Recio-Blanco}, {Piotto}, {Aparicio}, \& {Renzini}}]{Recio2002}
{Recio-Blanco}, A., {Piotto}, G., {Aparicio}, A., \& {Renzini}, A. 2002, \apjl, 572, L71

\bibitem[{{Recio-Blanco} {et~al.}(2004){Recio-Blanco}, {Piotto}, {Aparicio}, \& {Renzini}}]{Recio2004}
{Recio-Blanco}, A., {Piotto}, G., {Aparicio}, A., \& {Renzini}, A. 2004, \aap, 417, 597

\bibitem[{{Santos-Peral} {et~al.}(2021){Santos-Peral}, {Recio-Blanco}, {Kordopatis}, {Fern{\'a}ndez-Alvar}, \& {de Laverny}}]{Santos-Peral2021}
{Santos-Peral}, P., {Recio-Blanco}, A., {Kordopatis}, G., {Fern{\'a}ndez-Alvar}, E., \& {de Laverny}, P. 2021, \aap, 653, A85

\bibitem[{{Soubiran} {et~al.}(2022){Soubiran}, {Brouillet}, \& {Casamiquela}}]{Caroline22}
{Soubiran}, C., {Brouillet}, N., \& {Casamiquela}, L. 2022, \aap, 663, A4

\bibitem[{{Taylor}(2005)}]{Topcat}
{Taylor}, M.~B. 2005, in Astronomical Society of the Pacific Conference Series, Vol. 347, Astronomical Data Analysis Software and Systems XIV, ed. P.~{Shopbell}, M.~{Britton}, \& R.~{Ebert}, 29

\bibitem[{{Torres} {et~al.}(2006){Torres}, {Quast}, {da Silva}, {de La Reza}, {Melo}, \& {Sterzik}}]{Torres06}
{Torres}, C.~A.~O., {Quast}, G.~R., {da Silva}, L., {et~al.} 2006, \aap, 460, 695

\bibitem[{Weise {et~al.}(2010)Weise, Launhardt, Setiawan, \& Henning}]{Weise2010}
Weise, P., Launhardt, R., Setiawan, J., \& Henning, T. 2010, Astronomy \& Astrophysics, 517, A88

\bibitem[{{Wenger} {et~al.}(2000){Wenger}, {Ochsenbein}, {Egret}, {Dubois}, {Bonnarel}, {Borde}, {Genova}, {Jasniewicz}, {Lalo{\"e}}, {Lesteven}, \& {Monier}}]{Simbad}
{Wenger}, M., {Ochsenbein}, F., {Egret}, D., {et~al.} 2000, \aaps, 143, 9

\bibitem[{{Worley} {et~al.}(2016){Worley}, {de Laverny}, {Recio-Blanco}, {Hill}, \& {Bijaoui}}]{worley2016}
{Worley}, C.~C., {de Laverny}, P., {Recio-Blanco}, A., {Hill}, V., \& {Bijaoui}, A. 2016, \aap, 591, A81

\bibitem[{{Worley} {et~al.}(2012){Worley}, {de Laverny}, {Recio-Blanco}, {Hill}, {Bijaoui}, \& {Ordenovic}}]{worley2012}
{Worley}, C.~C., {de Laverny}, P., {Recio-Blanco}, A., {et~al.} 2012, \aap, 542, A48

\bibitem[{Zorec \& Royer(2012)}]{Zorec2012}
Zorec, J. \& Royer, F. 2012, Astronomy \& Astrophysics, 537, A120

\bibitem[{{Zorec} \& {Royer}(2012)}]{Zorec12}
{Zorec}, J. \& {Royer}, F. 2012, \aap, 537, A120

\bibitem[{{Z{\'u}{\~n}iga-Fern{\'a}ndez} {et~al.}(2021){Z{\'u}{\~n}iga-Fern{\'a}ndez}, {Bayo}, {Elliott}, {Zamora}, {Corval{\'a}n}, {Haubois}, {Corral-Santana}, {Olofsson}, {Hu{\'e}lamo}, {Sterzik}, {Torres}, {Quast}, \& {Melo}}]{Zuniga21}
{Z{\'u}{\~n}iga-Fern{\'a}ndez}, S., {Bayo}, A., {Elliott}, P., {et~al.} 2021, \aap, 645, A30

\end{thebibliography}

\end{document}